\documentclass[11pt,a4paper,preprintnumbers]{article}

\pdfoutput=1

\def\beq{\begin{equation}}
\def\eeq{\end{equation}}
\def\beqa{\begin{eqnarray}}
\def\eeqa{\end{eqnarray}}

\usepackage[pdftex]{graphicx}

\newcommand{\eqend}[1]{\,\mathrm{#1}}

\newcommand{\abs}[1]{{\left\vert{#1}\right\vert}}

\newcommand{\total}{\mathop{}\!\mathrm{d}}

\newcommand{\mathe}{\mathrm{e}}

\usepackage{jcappub}

\usepackage{amsmath}
\usepackage{amsfonts}
\usepackage{mathtools}
\usepackage{bm}

\subheader{\rm QGASLAB-14-01, KOBE-TH-14-01}

\title{Schwinger effect in de Sitter space
}
\author[a]{Markus B. Fr\"ob,}
\author[a]{Jaume Garriga,}
\author[b]{Sugumi Kanno,}
\author[c]{Misao Sasaki,}
\author[d]{Jiro Soda,}
\author[c]{Takahiro Tanaka,}
\author[e] {Alexander Vilenkin}
\affiliation[a]{Departament de F{\'\i}sica Fonamental i \\
Institut de Ci{\`e}ncies del Cosmos, 
Universitat de Barcelona,\\
Mart{\'\i}\ i Franqu{\`e}s 1, 08028 Barcelona, Spain}
\affiliation[b]{Laboratory for Quantum Gravity \& Strings and Astrophysics, Cosmology \& Gravity Center,
Department of Mathematics \& Applied Mathematics, University of Cape Town,
Private Bag, Rondebosch 7701, South Africa}
\affiliation[c]{ Yukawa Institute for Theoretical Physics, Kyoto University,
Kyoto 606-8502, Japan}
\affiliation[d]{Department of Physics, Kobe University, Kobe 657-8501, Japan}
\affiliation[e]{Institute of Cosmology, Department of Physics and Astronomy, 
Tufts University, Medford, Massachusetts 02155, USA}

\abstract{We consider Schwinger pair production in 1+1 dimensional de Sitter space, filled with a constant electric field $E$.
This can be thought of as a model for describing false vacuum decay beyond the semiclassical approximation, where pairs of a quantum field $\phi$ of mass $m$ and charge $e$ play the role of vacuum bubbles.  We find that the adiabatic ``in" vacuum associated with the flat chart 
develops a space-like expectation value for the current $J$, which manifestly breaks the de Sitter invariance of the background fields.  We derive a simple expression for $J(E)$, showing that both ``upward" and ``downward" tunneling contribute to the build-up of the current. For heavy fields, with $m^2\gg eE,H^2$, the current is exponentially suppressed, in agreement with the results of semiclassical instanton methods. Here, $H$ is the inverse de Sitter radius. On the other hand, light fields with $ m \ll H$ lead to a phenomenon of infrared
hyperconductivity, where a very small electric field $mH \lesssim eE \ll H^2$
leads to a very large current $J \sim H^3 /E$.
We also show that all Hadamard states for $\phi$ necessarily break de Sitter invariance. 
Finally, we comment on the role of initial conditions, and ``persistence of memory" effects. 
}

\begin{document}
\maketitle

\section{Introduction}

False vacuum decay through bubble nucleation \cite{VKO,Coleman,CdL,Linde} has received considerable attention in field theory and cosmology over the past few decades. Interest in this subject has recently been revived in the context of the eternally inflating multiverse \cite{GSVW,Carroll,BSV}. Bubble nucleation may play an important role in determining the large scale structure of the multiverse and the distribution of vacua within it. Various suggestions have  also been made for possible observational signatures of this scenario, involving the dynamics of bubble formation \cite{BS,S,KS} or subsequent bubble collisions \cite{GGV,KLS,SSS,WJPALL}.

Although the mechanism for vacuum decay by quantum tunneling seems to be reasonably well understood, some aspects of it require further exploration.
A particularly puzzling issue which has only recently been addressed \cite{GKSSV,GKT}, concerns the rest frame of bubble nucleation. If the false vacuum is locally Lorentz invariant, what is it that determines the frame in which bubbles of the new vacuum nucleate at rest? In principle, we may expect this to be partially determined by the hypersurface of initial conditions, where the false vacuum is prepared, and partially by the state of motion of the detectors which should be used in order to probe the process
of bubble formation.

A convenient framework for investigating this question is Schwinger pair production in 1+1 dimensions. In this setup, the nucleated pairs can be treated fully quantum mechanically, and not just semiclassically as is customary with vacuum bubbles.  Refs. \cite{GKSSV,GKT} concentrated on the case of a constant electric field in flat space. In this case, it was shown that the adiabatic in-vacuum for a charged scalar field $\phi$ 
(defined in terms of modes which are positive frequency in the remote past) is Lorentz invariant (LI). Then, by using various ÒdetectorÓ models, it was shown that particles and antiparticles tend to nucleate preferentially at rest in the detector's frame. 

On the other hand, the Lorentz invariant vacuum corresponds to a somewhat idealized situation which is not too realistic. If the electric field is switched on at some initial time $t_0$, then in the limit $t_0\to -\infty$ the number of pairs which have been produced per unit volume is infinite for any finite value of $t$. Because of that, the LI vacuum contains an infinite density of charged particles, the two point function for $\phi$ does not have the Hadamard form, and the expectation value of the current is ill defined \cite{GKSSV}. 
For a more realistic case, where we keep $t_0$ finite, the number density of produced pairs is finite, leading to a space-like expectation value for the current of the form 
$J^{\mu} = (0,J)$, with (see, e.g., \cite{Mottola1,Mottola2} and references therein):
\begin{equation}
J\approx 2 e (t-t_0) \Gamma. \label{flatcurrent}
\end{equation}
Here, $e$ is the electric charge and 
\begin{equation}
\Gamma={eE\over 2\pi} e^{-\pi{m^2\over eE}}\label{srate}
\end{equation}
 is the pair production rate per unit volume (see e.g. Ref.\cite{Cohen} and references therein), where $m$ is the mass of the charge carrier and $E$ is the electric field.\footnote{The time it takes for a given pair to be excited out of the vacuum can be estimated as \cite{GKT} $\tau_{nuc} \sim r_0$, where
$r_0$ is the size of the instanton, given in (\ref{size}). Hence, in (\ref{flatcurrent}) we also assume $t-t_0 \gg \tau_{nuc} \sim r_0.$}
In principle, the breaking of Lorentz invariance by the initial hypersurface at $t=t_0$ could have some influence on the frame
of nucleation. However, it was argued in \cite{GKT} that such influence becomes irrelevant in the asymptotic future, when the proper time
$\tau$ which the detector has spent in the false vacuum exceeds the size $r_0$ of the instanton which contributes to the decay rate
\begin{equation}
\tau\gg r_0 = {m\over eE}. \label{size}
\end{equation}
Pairs would then nucleate in the detector's rest frame to very good approximation, essentially as if the system were in the Lorentz invariant vacuum.

In this paper, we shall study the Schwinger effect in de Sitter space (dS), which is more relevant to the inflationary context. One of  our goals will be to clarify the role of initial conditions. As we shall see, in the presence of an electric field, all Hadamard vacua for charged particles have the property that they break dS invariance.  The symmetry breaking can be attributed to initial conditions, whose influence persists for arbitrarily late times. This is related to the ``persistence of memory" effect first discussed in Ref. \cite{GGV}.

The Schwinger process in 1+1 dS space has previously been considered in \cite{j}. The distribution in the number density of particles created by the electric and gravitational fields was calculated by using the method of Bogoliubov coefficients, and it was shown that in the semiclassical limit the result agrees with instanton computations \cite{j2}. This applies both to ``downward" tunneling, 
where the initial false vacuum is more energetic than the new vacuum, and to ``upward" tunneling, where the new  vacuum is more energetic than the initial one. 
Upward tunneling \cite{LW} is possible during inflation because energy is not conserved on scales larger than the horizon size. This is relevant in determining the frequency at which different vacua in the landscape are visited by a hypothetical observer as a result of vacuum transitions. Perfect ergodicity in the frequency of such visits would imply the absence of a thermodynamic arrow of time, and so the precise rate of upward transitions seems to be important  at a fundamental level (see Ref. \cite{watcher} for a recent discussion of this issue). Since a rigorous justification of instanton methods in dS is still lacking, particularly for upward tunneling, the results of \cite{j} provide valuable evidence for the quantitative accuracy of this approach. 

The calculation of particle creation done in \cite{j} is based on the existence of an adiabatic ``out" vacuum in the asymptotic future. This, in turn, requires the mass of the particles to be much larger 
than the Hubble rate $m\gg H$. We can go beyond this regime by considering the expectation value of the current, $J(E)$, which is generated by the electric field as a result of the Schwinger process. As we shall see, this observable receives distinct contributions both from upward and downward tunneling, and it is well defined regardless of the existence of adiabatic asymptotic regions. 
Besides, the investigation of the behaviour of the current as a function of the applied electric field seems worth pursuing in its own right, and we shall see that the vacuum shows an interesting phenomenon of infrared hyperconductivity (with possible relevance for cosmology).

We start in Section \ref{schsec}, by reviewing pair production by an electric field in 1+1 dimensional dS. The semiclassical limit,
relevant for comparison  with bubble nucleation, is discussed in Section \ref{time}. Based on the semiclassical picture, we give a heuristic 
derivation of the current in Section \ref{analytic}. Surprisingly, this coincides with the exact result for the renormalized expectation value of the current, which is calculated in Section \ref{currentsec}. 
The non-vanishing expectation value of the current in the ``in" vacuum manifestly breaks dS invariance. 
Thisis in contrast with the case of flat space, where, as mentioned above, the adiabatic 
``in" vacuum is  Poincar\'e invariant (and non-Hadamard), and the expectation value of the current is ill defined. 
We then analyze the conductivity of vacuum in different regimes, characterized by the mass $m$ of the charge carriers and the strength of the electric field $E$.  
In Section \ref{hadamard} we consider the question of dS invariance in more general terms, showing that it is broken in any Hadamard vacuum. 
In Section \ref{persistence} we discuss some aspects of the persistence of memory of initial conditions. We argue that, unlike in the case of flat space, the influence of the initial hypersurface in determining the rest frame of nucleation is a persistent feature in de Sitter space. Our conclusions are summarized in Section \ref{conclusions}. Appendix \ref{app} discusses the semiclassical trajectories of charged particles in dS, Appendix \ref{integral} contains the calculation of the current from a momentum integral of special functions, and Appendix \ref{detector} deals with particle detectors.

\section{Schwinger effect in dS}\label{schsec}

Consider a 1+1 dimensional de Sitter space with a constant electric field $E$. The field strength is given by
\begin{equation}
F_{\mu\nu} = -E  \sqrt{-g} \epsilon_{\mu\nu}.\label{fs}
\end{equation}
Here $\epsilon_{\mu\nu}$ is the Levi-Civita symbol, with $\epsilon_{01} =1$, and $g$ is the determinant of the dS metric $g_{\mu\nu}$. The symmetry 
of this background is SO(2,1), rather than the full de Sitter group O(2,1), since the field strength is not invariant under parity. This distinction, however, 
is not too relevant for our purposes, and for brevity we shall simply refer to SO(2,1) symmetry as dS invariance. 
An important feature of this background is that it does not possess any preferred rest frame.

Following \cite{j},  let us consider a charged scalar field $\phi$ with action given by
\begin{equation}
S=\int d^2 x \sqrt{-g}\left[-g^{\mu\nu}(\partial_\mu+i e A_\mu)\phi^*(\partial_\nu-ieA_\nu)\phi - m^2 \phi^*\phi \right].
\label{action}
\end{equation}
In the flat chart, the dS metric reads
\begin{equation}
ds^2 = {1\over (H\eta)^2} [-d\eta^2 + dx^2], \quad\quad (-\infty<\eta<0) \label{metric}
\end{equation}
and the gauge potential leading to (\ref{fs}) can be taken as 
\begin{equation}
A_\mu = {E \over H^2 \eta} \delta^x_\mu.   \label{gaugepot}
\end{equation}
Owing to the spatial homogeneity of  (\ref{metric}) and (\ref{gaugepot}), we can expand the field operator as 
\begin{equation}
\phi = \int_{-\infty}^\infty {dk\over (2\pi)^{1/2}} \left(a_k \phi_k(\eta) + b^\dagger_{-k} \phi^*_k(\eta)\right) e^{ikx}, \label{fieldexp}
\end{equation}
and then the equation of motion reduces to
\begin{equation}
\phi_k'' + \left[ {m^2\over H^2\eta^2} + \left(k-{eE\over H^2 \eta}\right)^2\right] \phi_k =0, \label{meq}
\end{equation}
where primes indicate derivatives with respect to $\eta$.
The canonical commutation relations imply $[a_k,a^\dagger_{k'}]=\delta(k-k')$, $[b_k,b^\dagger_{k'}]=\delta(k-k')$, with the modes $\phi_k$ satisfying the standard Klein-Gordon normalization condition
\begin{equation}
i(\phi^*_k\phi'_k - \phi_k\phi^{*'}_k) =1.
\end{equation}
The ``in" vacuum corresponds to the choice of modes
\begin{eqnarray}
\phi_k^{\rm in} &=& (2 k)^{-1/2} e^{-\pi|\lambda|/2} W_{\lambda,\sigma} (2ik\eta),\quad(k>0) \label{modes0} \\
\phi_k^{\rm in}&=& (2|k|)^{-1/2} e^{\pi |\lambda|/2}  W_{-\lambda,\sigma} (2i|k|\eta),\quad(k<0)   \label{modes}
\end{eqnarray}
where $W_{\lambda,\sigma}$ are Whittaker functions, with indices given by $\lambda = i eE/H^2$, and
\begin{equation}
\sigma=\left({1\over 4} - {e^2E^2\over H^4}-{m^2 \over H^2}\right)^{1/2}. \label{sigma}
\end{equation}
In the case of heavy particles, for which $m^2\gg H^2$, $\sigma$ is purely imaginary. In this case we adopt the convention\footnote{Note that $W_{\lambda,\sigma}(z) = W_{\lambda,-\sigma}(z)$, and so the ambiguity in the sign of the square root in the right hand side of 
(\ref{sigma}) is irrelevant.}
\begin{equation}
\sigma = i|\sigma|.
\end{equation}
From the asymptotic expansion of $W_{\lambda,\sigma}(z) \sim e^{-z/2} z^{\lambda}$ for large $|z|$, the modes (\ref{modes}) behave as
\begin{equation}
\phi_k^{\rm in} \sim {1\over\sqrt{2|k|}} e^{-i|k|\eta}  (-2|k|\eta)^{i|\lambda| {\rm sign}(k)} \{1+O[(k\eta)^{-1}]\}, \quad (|k|\eta\to -\infty) \label{asin}
\end{equation}
and so they are positive frequency with respect to conformal time in the asymptotic past. 
Such ``in" vacuum can be used in order to calculate pair production rates by the method of Bogoliubov coefficients. The ``out'' vacuum can be defined (for heavy particles with $m\gg H$) by using the Whittaker function 
$M_{\lambda,\sigma}$, in terms of which the mode functions are given by
\begin{equation}
\phi_k^{\rm out} = {1\over \sqrt{2|k\sigma|}} e^{-{\pi\over 2}|\sigma| {\rm sign}(k)} M_{\lambda,\sigma}(2ik\eta),
\end{equation} 
For $|z|\ll 1$, we have $M_{\lambda,\sigma}(z) \sim z^{\sigma+{1\over 2}}[1+O(z)]$, and so in the asymptotic future we have
\begin{equation}
\phi_k^{\rm out} \sim {1\over \sqrt{2H|\sigma|}} e^{-i H|\sigma| t} e^{-{Ht\over 2}} [1+O(kH^{-1} e^{-Ht})], \quad (|k|\eta\to 0^-)
\end{equation}
which is positive frequency with respect to cosmological time $t$. This is related to conformal time through
\begin{equation}
a(t)=e^{Ht}=-(H\eta)^{-1}. \label{prot}
\end{equation}
The ``in" and ``out" modes are related by the Bogoliubov coefficients $\alpha_k$ and $\beta_k$:
\begin{equation}
\phi_k^{\rm in}=\alpha_k\ \phi_k^{\rm out} +\beta_k\ \phi^{{\rm out}*}_k.
\end{equation}
These can be read off from the linear relation
\begin{equation}
W_{\lambda,\sigma}(z) = {\Gamma(-2\sigma)\over \Gamma\left({1\over 2}-\sigma-\lambda\right)} M_{\lambda,\sigma}(z) + {\Gamma(2\sigma)
\over \Gamma\left({1\over 2}+\sigma-\lambda\right)} M_{\lambda,-\sigma}(z),
\end{equation}
from which we easily obtain  \cite{j}
\begin{equation}
|\beta_k|^2 = |\beta_{\pm}|^2 = e^{-\pi(|\sigma|\pm|\lambda|)}{\cosh\pi(|\sigma|\mp|\lambda|) \over \sinh 2\pi|\sigma|}.\label{bogo}
\end{equation}
Here, the upper sign corresponds to $k>0$, and the lower sign to $k<0$.

In flat space, an electric field causes particles and antiparticles in a pair to nucleate  at a distance $d=2 r_0$ from each other. The distance $d$ is determined by the balance between the potential energy and rest mass energy $eE d=2m$, where $e$ is the electric charge. If $E>0$, this balance requires that the particle with positive charge should be to the right of the particle with negative charge (i.e., towards increasing values of $x$). We may call this the ``screening" orientation, since the charges would then tend to reduce the value of the electric field in between them. In the language of false vacuum decay, this corresponds to a ``downward" transition, reducing the value of the vacuum energy density. Here, we shall treat the electric field as an external source, which will be unaffected by the nucleation of pairs, but we shall still refer to the materialization of pairs with the screening orientation as ``downward" tunneling. 
In de Sitter space, pairs can also nucleate with the ``anti-screening" orientation, since energy need not be conserved on scales somewhat bigger than $H^{-1}$. This corresponds to ``upward" tunneling \cite{LW,recycling}. Fig. 1 illustrates the semiclassical trajectories of two nucleating pairs. 
Downward tunneling corresponds to the excitation of modes with $k<0$, while upward tunneling corresponds to the excitation of modes with $k>0$.

\begin{figure}
\begin{center}
\vspace{-2cm}
\includegraphics[width=14cm]{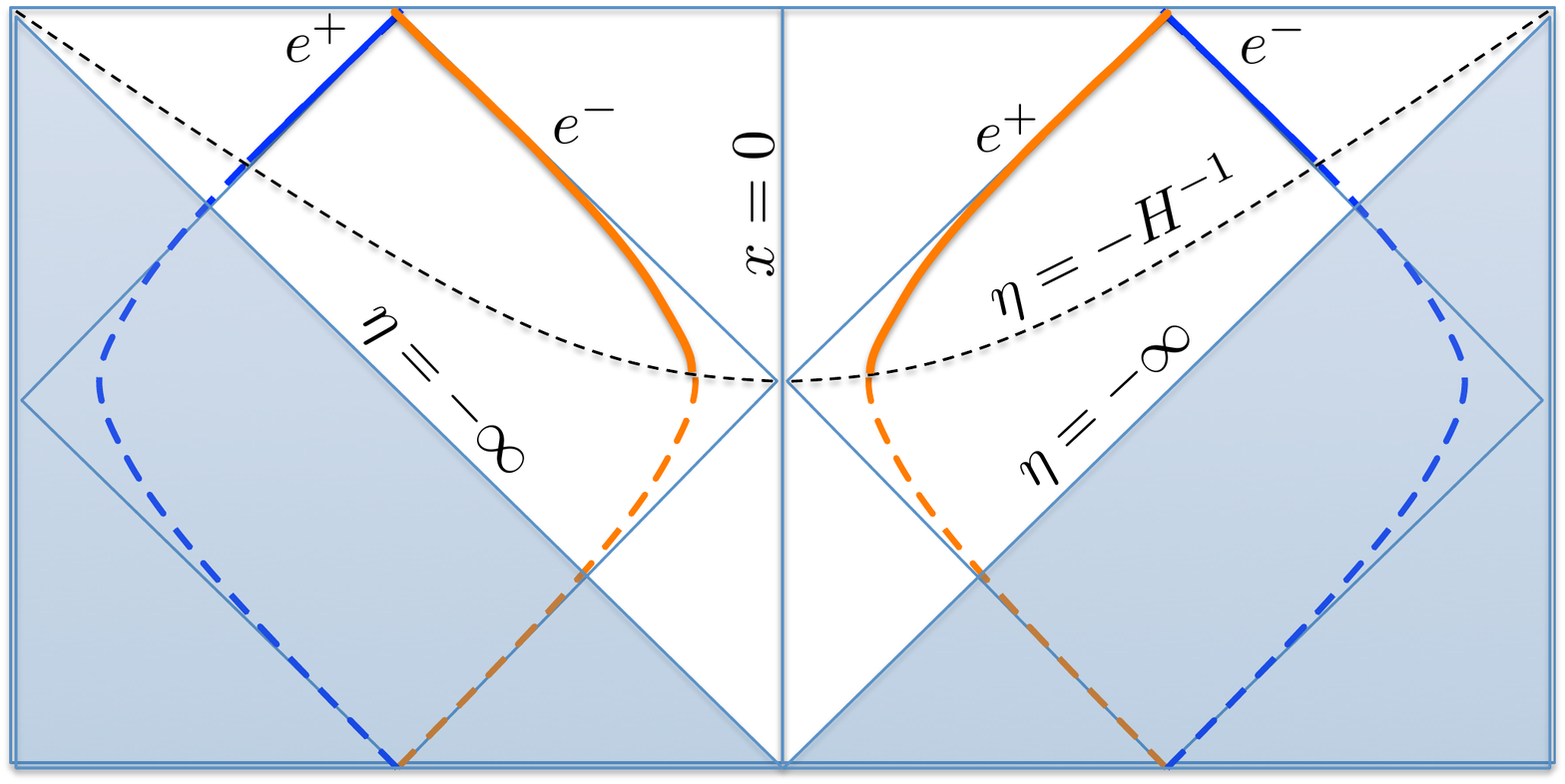}
\vspace{-1.5cm}
\caption{Diagram illustrating the nucleation of charged pairs in a 1+1 dimensional de Sitter space with a constant electric field $E$. The white region corresponds to the patch which is covered by the flat chart, with
coordinates $(\eta,x)$. The adiabatic ``in" state in the flat chart does not contain any particles at $\eta\to -\infty$, which can be thought of as the hypersurface of initial conditions. The shaded part of the diagram is 
irrelevant for our discussion. If the electric field $E>0$ points in the positive $x$ direction, pairs can nucleate with the usual ``screening" orientation (red) or the ``antiscreening" orientation (blue). The former corresponds to downward tunneling, and the latter to upward tunneling.}
 \label{fig1}
\end{center}
\end{figure}

\section{Semiclassical limit} \label{time}

We may now elaborate on the semiclassical description of pair creation. For 
\begin{equation}
|\sigma|\pm|\lambda| \gg 1, \label{scrange}
\end{equation}
Eq. (\ref{bogo}) gives 
\begin{equation}
|\beta_\pm|^2 \approx e^{-2\pi(|\sigma|\pm|\lambda|)}. \label{scb}
\end{equation}
In flat space, pair creation is entirely due to the electric field, but in an expanding background, such as dS,  pairs can be produced even if the electric field vanishes. In order to 
characterize the relative importance of these two effects, we introduce the parameter 
\begin{equation}
\ell \equiv {eE\over \tilde m H}
\end{equation}
where, 
\begin{equation} 
\tilde m^2 = m^2 -{H^2 \over 4}.
\end{equation}
In this notation, Eq. (\ref{scb}) reads
\begin{equation}
|\beta_\pm|^2 \approx e^{-S_{\pm}} \equiv \exp\left[-{2\pi \tilde m\over H} \left(\sqrt{1+\ell^2}\pm\ell\right)\right].\label{scs}
\end{equation} 
In this Section we shall only be concerned with the semiclassical limit, where $S_\pm$ is large. A necessary condition is that the mass be large 
compared with $H$. In this regime $\ell$ will be real.

For $\ell \ll 1$ pairs are mainly produced by the cosmological expansion, and we have
\begin{equation}
S_\pm \approx 2\pi\left({\tilde m \over H} \pm {eE\over H^2} \right), \quad \quad (\ell \ll 1).
\end{equation}
The first term corresponds to the Boltzmann factor for non-relativistic massive particles at the Gibbons-Hawking temperature, while the second can be though of as a small correction due to the electric field.  In the opposite limit, $\ell \gg 1$, we have 
\begin{eqnarray}
S_+ &\approx& 4\pi {eE \over H^2}, \quad \quad (\ell\gg 1), \label{upth}\\
S_- &\approx& \pi {\tilde m^2 \over e E},       \quad\quad (\ell\gg1). \label{schwexp}
\end{eqnarray}
The result (\ref{upth}) for $S_+$ corresponds to upward tunneling, where the separation of the particles in a pair at the time of production is comparable to the horizon size, while $S_-$ reduces to the standard semiclassical instanton action for the Schwinger process in  flat space. Note that $S_+ \sim \ell^2  S_-\gg S_-$, so upward tunneling is highly suppressed compared to downward tunneling in this limit.

In order to estimate the time at which the pairs in a given mode $k$ are excited out of the vacuum, we may adopt the criterion that this occurs
when the violation of adiabaticity in the corresponding mode is maximal. To analyze this issue, it is convenient to introduce
 \begin{equation}
 \psi_k=a^{1/2} \phi_k,
 \end{equation}
The mode equation (\ref{meq}) can now be rewritten as
 \begin{equation}
 \ddot \psi_k + w_k^2\ \psi_k =0.
 \end{equation} 
Here dots indicate derivative with respect to proper time $t$, defined in (\ref{prot}), and
 \begin{equation}
w_k^2 \equiv \tilde  m^2[1 + \ell^2 (z+1)^2], \label{wk}
 \end{equation} 
with
 \begin{equation}
 z= {H\over eE }\left({k\over a}\right).\label{zdef}
  \end{equation}
The frequencies $w_k$ approach a constant in the asymptotic future, leading to a well defined notion of ``out" particles.
Let us now show that the adiabatic condition  
 \begin{equation}
 f_k(z) \equiv \left|{\dot w_k \over w_k^2}\right| = \left({H\ell^2\over \tilde m}\right){|(z+1) z| \over [\ell^2 (z+1)^2+1]^{3/2}} \ll 1,\label{acon}
 \end{equation}
 is well satisfied (at all times) in the semiclassical parameter range given by (\ref{scrange}). Tiny deviations from perfect adiabaticity will lead to the exponentially
 suppressed expectation values (\ref{scb}) for the out particle numbers. 
 
 The violation of adiabaticity is largest at the extrema of $f_k$. From $d\log f_k/dz=0$, we have
\begin{equation}
\ell^2(z^2-1) = 2-{1\over z+1}. \label{intsec}
\end{equation}
which can be seen as the intersection of a parabola with a hyperbola. This has three real solutions for $z$. 
For $\ell\gg1$, these are given by 
\begin{eqnarray}
z&=&z_- \approx -1\pm (\sqrt{2}\ell)^{-1}, \quad\quad (k<0),  \label{zm} \\ 
z&=&z_+\approx 1+(3/4)\ell^{-2}, \quad\quad (k>0).  \label{zp}
\end{eqnarray}
The subindices in $z_\pm$ refer to the fact that, according to (\ref{zdef}),  the sign of $z$ coincides with the sign of $k$.
The corresponding maximum values of the adiabaticity parameters are given by 
\begin{equation}
f_\pm \sim 1/S_{\pm} \ll1,
\end{equation}
Therefore $f_k\leq \max(f_\pm) \ll 1$, as advertised in (\ref{acon}).
Particle creation in mode $k$ occurs around the time $t_k$ when $f_k$ is maximum. Using (\ref{zm}-\ref{zp}) in (\ref{zdef}), 
we are led to the estimate
\begin{equation}
{k\over a(t_k)} \approx \pm  {eE\over H}, \quad\quad (\ell\gg1).\label{tkg}
\end{equation}
For $\ell \ll 1$ two of the roots of (\ref{intsec}) are given by
\begin{equation}
z=z_\pm \approx \pm\sqrt{2} \ell^{-1}, \label{zpm}
\end{equation}
also with $f_\pm \sim 1/S_\pm \ll 1$. There is a third root at 
$z=z_3\approx -1/2$, which is negative just like $z\approx z_-$. This is also an extremum of $f_k(t)$ for $k<0$. However, the adiabaticity parameter $f_3=f_k(z_3)$ is suppressed with respect to $f_-$ by a factor of $\ell^2$. Hence, for $k<0$ the main departure from adiabaticity occurs at $z\approx z_-$. Using (\ref{zpm}) in (\ref{zdef}),
we find that the time of particle creation is given by
\begin{equation}
{k\over a(t_k)} \approx \pm \sqrt{2}\ \tilde m, \quad\quad (\ell\ll1).\label{tks}
\end{equation}
Eqs. (\ref{tkg}) and (\ref{tks}) can be compressed in the following estimate
\footnote{ The width of the peaks of the adiabaticity parameter can be estimated by calculating the second derivatives of $f$. 
The two peaks given in (\ref{zp}) and (\ref{zpm}) have 
widths of order $(\Delta z/ z) \sim S_\pm^{-1/2}\ll 1$, so they are very sharp in the semiclassical limit. On the other hand, the double peaks given in Eq. (\ref{zm})
have a width  $(\Delta z/ z) \sim \ell^{-1} S_\pm^{-1/2}\ll 1$. Here, however, the important parameter is not so much the width of the individual peaks but the separation between them. This is
given by $\Delta  z \sim \ell^{-1}$, which corresponds to a time difference $\Delta t \sim r_0$, where $r_0 =m/(eE)$ is the instanton radius.
This is in concordance with the case of flat space \cite{GKT}.\label{width}} 
\begin{equation}
{k\over a(t_k)} \sim \pm |\sigma| H, \label{tk}
\end{equation}
for the time $t_k$ at which pair creation occurs in mode $k$ \cite{j}.

The number distribution of created pairs per unit co-moving volume is given by
\begin{equation}
{dN\over dx} = |\beta_k|^2 {dk\over 2\pi}. \label{distribution}
\end{equation}
Since $|\beta_k|$ depends only on the sign of $k$, the distribution is flat both for positive and negative $k$, but discontinuous at $k=0$. 
Also, at any finite value of $t$, the 
distribution is cut-off at $|k| \sim a(t) |\sigma| H$, since according to (\ref{tk}), modes with a higher value of $|k|$ have not yet been excited.
The distribution (\ref{distribution}) can be compared with the distribution which is obtained by means of instanton techniques \cite{j2}. The use of instanton methods in dS is not as rigorously justified as it is in flat space. Nonetheless, 
it was found in \cite{j,j2} that the results of instanton and Bogoliubov methods agree in the semiclassical limit (i.e., when $|\sigma|\pm |\lambda|\gg 1$) 
not just in the exponential dependence, but also in the one loop prefactor. This result applies both to downward ($k<0$), and upward transitionsÊ ($k>0$). 

For a charged particle of momentum $k$, the physical momentum with respect to the co-moving observers is given by
\begin{equation}
p=a^{-1}(k-eA_x)= {k\over a} + {eE \over H}.\label{pk}
\end{equation}
Note that all particles approach a terminal value of the physical momentum at late times $a\to \infty$,
\begin{equation}
p_{\infty} = {eE \over H}, 
\end{equation}
which is positive for particles and negative for antiparticles. This can be interpreted as the momentum which is gained by a charged particle subject to a constant electric field during a Hubble time. Additional time does not increase the physical momentum relative to the co-moving frame, since momentum is also depleted due to Hubble friction. 
A particle with $k<0$ has $p<0$ at early times and $p>0$ at late times, with a turning point at 
\begin{equation}
{k\over a(t_{turn})}=- p_\infty =-{eE\over H}. \label{turning}
\end{equation} 
For such particles, the terminal velocity is approached from below, so $|p|<p_\infty$ at all times. A particle with $k>0$ always has $p>p_\infty>0$, and the terminal velocity is approached from above, without any turning points (see Fig. \ref{fig1}). 

Finally, let us comment on a puzzling aspect concerning the time of pair creation. For large electric field ($\ell \gg 1$) and downward tunneling ($k<0$), according to Eqs. (\ref{tkg}) and (\ref{turning}), pairs are produced at the turning point in the semiclassical trajectory. This is in agreement with the situation in flat space. On the other hand, for small (or even vanishing) electric field ($\ell \ll 1$), heavy particles with $m^2\gg H^2$ have a sizeable physical momentum $p_c$ at the time of creation. Using (\ref{tks}), we have
\begin{equation}
p_c^2\sim m^2. \label{relmom} 
\end{equation}
This seems to be at odds with the fact that, in the absence of the electric field, an inertial Unruh 
detector coupled to $\phi$ will reach thermal equilibriun at the temperature $T=(2\pi)^{-1} H$, as if it were immersed in a thermal bath.
Note that a true thermal bath of heavy particles in flat space has a root mean squared value of the momentum given by
\begin{equation}
\langle p^2\rangle_T \sim m T \ll m^2,
\end{equation}
which is much smaller than (\ref{relmom}). A related observation is that the momentum distribution of $\phi$ particles, given by (\ref{distribution}), is not thermal at all. Rather, as mentioned above, it is completely flat, with a cut-off at the physical momentum of order $p_c$. Nonetheless, as we shall see in the following Sections, if instead of using an Unruh
detector we use an amperemeter that measures the average current flowing in response to a small electric field, the result is consistent with
a flat distribution of the form (\ref{distribution}), with a cut-off of the form (\ref{tk}).


  \section{Semiclassical current} \label{analytic}
 
 In this Section, we give a heuristic derivation of the current based on the semiclassical picture. 
 The current due to semiclassical particles after pair creation is given by
\begin{equation}
J_{\rm pairs}= 2e \int v\ dn,
\end{equation}
where $dn= a^{-1} dN/dx=|\beta_k|^2 dk/(2\pi a)$ is the diferential number density of carriers and $v$ is their velocity.
Separating this into two components, $J_{\rm pairs}=J^+_{\rm pairs}+ J^-_{\rm pairs}$, corresponding to $k>0$ and $k<0$ respectively, we have
\begin{equation}
J^\pm_{\rm pairs}= {e\over\pi a}\int_0^{k_c^\pm} {dk} {p^{\pm} \over \sqrt{m^2+(p^{\pm})^2}} |\beta^\pm|^2, \label{curint}
\end{equation}
The physical momentum is given by $p^{\pm}=\pm(k/a)+|\lambda|H$, and 
the upper limit of integration is taken from 
(\ref{tk}),
\begin{equation}
{k_c^\pm} \sim a(t) |\sigma| H. \label{tk2}
\end{equation}
Here we are using the notation $k_c^{\pm}$ to denote the absolute value $|k_c|$ of the cut-off momentum, which in principle can be different for upward or downward tunneling. The uncertainty in the cutoff is of the order of the width of the peak in the adiabaticity parameter\footnote{See footnote \ref{width}.}. We do not need to be too precise about the value of the momentum cutoff $k_c$, but it will be important to know that it scales with $a(t)$.

Performing the integral (\ref{curint}), we have
\begin{equation}
J^{\pm}_{\rm pairs} = \pm{e\over\pi} \left(m\gamma_c^\pm - |\sigma| H\right) |\beta^\pm|^2. \label{curint2}
\end{equation}
Here, $\gamma_c^\pm$ stands for the relativistic gamma factor, $\gamma= (1+p^2/m^2)^{1/2}$ , evaluated at the cut-off values of the momentum. 

If we take equal values for the momentum cutoff $k_c^-=k_c^+$, then the current due to semiclassical particles takes the form
\begin{equation}
J_{\rm pairs} = J^+_{\rm pairs} + J^-_{\rm pairs} = {e \over\pi} \left(m\gamma_c- |\sigma| H\right) \left(|\beta^+|^2-|\beta^-|^2\right).
\label{jpairs}
\end{equation}
For $\ell \ll 1$ we have $|\lambda|\ll |\sigma|$ and
\begin{equation}
|\beta^\pm|^2\approx e^{-S_\pm} \approx e^{-2\pi m/H} e^{\mp 2\pi |\lambda|}. \label{betas} 
\end{equation}
Note that, since downward tunneling is more likely than upward tunneling, $|\beta^-|^2 > |\beta^+|^2$, the current due to semiclassical pairs (\ref{jpairs}) actually runs opposite to the electric field, which is somewhat counterintuitive.

On the other hand, we should take into consideration that the total semiclassical current is the sum of two contributions
\begin{equation}
J =J_{\rm pairs}+J_{\rm vac}, \label{totalj}
\end{equation}
where $J_{\rm vac}$ is the vacuum current which links the two members of a pair as they are created out of the vacuum. This is a space-like current which  is necessary for local charge conservation, and can be visualized as a line (which may perhaps be rather thick) connecting the negative charge with the positive charge at the moment of creation. For any given pair, which we may label with an index $i$, the vacuum current can be written as
\begin{equation}
J_{{\rm vac},i}^\mu = e \int ds\ {dx^{\mu} \over ds}\ {1\over \sqrt{-g}} \delta^{(2)}\left(x^\nu-x_i^\nu(s)\right).
\end{equation} 
Here, $x_i^{\mu}(s)$ parametrizes the locus where the current is non-vanishing, which for simplicity we take to be one-dimensional thin line.\footnote{The current should run inside of the flat chart, without taking a 
shortcut across its past boundary.}
As a crude approximation, we can take $x_i^\mu(s)$ to be on a $t=t_i=const.$ line, where $t_i$ is the moment when the $i$-th pair is created. In this case, we have
\begin{equation}
J_{\rm vac}(t) \equiv {1 \over V_t} \sum_{i\in V_t} a(t_i)  \int_{V_t} d^2 x \sqrt{-g} J_{{\rm vac},i}^x =  {e\over V_t} \sum_{i\in V_t} a(t_i) 
(\Delta x)_i.
\end{equation}
Here, $V_t$ indicates a 2-volume of infinitesimal thickness $\Delta t$ in the temporal direction, and arbitrarily large extent in the spatial direction $x$, and $(\Delta x)_i$ is the spatial coordinate separation between the positive and negative charges in the pair.
Note that 
\begin{equation}
d_i \equiv \mp a(t_i) (\Delta x)_i,
\end{equation}
is just the physical distance between the particle and antiparticle in the pair. Since the electric field and the Hubble rate are constant, this physical distance will be the same for all pairs of the same kind, and we immediately find
\begin{equation}
J_{\rm vac} = {e\over a(t)} \left(-d_+ {dN_+ \over dt dx} + d_- {dN_-\over dt dx}\right)= e \sum_\pm \mp {H d_\pm  \over 2\pi}{k_c^\pm \over a} |\beta^\pm|^2. \label{curvac}
\end{equation}
Here $N_\pm$ are the number of pairs with the anti-screening or screening orientation, given by
\begin{equation}
N_\pm = \int_0^{k_c^{\pm}} {dk \over 2\pi} |\beta_\pm|^2,
\end{equation}
and in the last step we have used that $k_c^\pm \propto a(t)$.

We show in Appendix \ref{app} that on the semiclassical trajectory, the following relation holds:
\begin{equation}
H d_\pm {k_c^\pm \over a(t)} = 2 m \gamma_c^\pm. \label{distance}
\end{equation}
Using this equation in (\ref{curvac}), and substituting the result in (\ref{totalj}), with $J_{\rm pairs}$ given by (\ref{curint2}), we have 
\begin{equation}
J = e{H |\sigma| \over \pi} \left(|\beta^-|^2 - |\beta^+|^2\right).  \label{janalytic}
\end{equation}
This expression explicitly shows the two distinct contributions from upward and downward tunneling (which are comparable for $|\lambda|\ll 1$).
It should be noted that in order to derive (\ref{janalytic}) we did not need to specify the precise cutoff values of $k_c^{\pm}$, but only had to assume that the cutoff of the flat distribution (\ref{distribution}) is at a fixed value of $k/a$. This is, of course, consistent with the estimate (\ref{tk}) for the time of pair creation, which was based on the analysis of the peak in the adiabaticity parameter. It is nice that the result for the current is robust against the uncertainties in the location of this peak, but this also means that this observable carries little information about the value of the momentum of the particles at the time of nucleation. Let us now compare the semiclassical expression (\ref{janalytic}) to the quantum expectation value of the current. As we shall see, the agreement turns out to be impressive.

\section{Expectation value of the current in the ``in" vacuum}\label{currentsec}

It was pointed out in \cite{j} that the flat chart ``in" vacuum is Hadamard. What is meant by this is that in the coincidence limit the two point function has the same divergences as a neutral field in the Bunch-Davies vacuum (BD), while it is finite when the two points are separated. In 1+1 dimensions, the divergence is actually the same as the logarithmic divergence in flat space. 
Let us first review the argument showing that the state is Hadamard. 
For later use, we introduce the gauge invariant two point functions \cite{GKSSV}
\begin{eqnarray}
G^+(x^\mu,y^\mu) = \langle \phi^\dagger(x^\mu)  e^{-ie\int_x^{y} A_\mu dx'^\mu}   \phi(y^\mu)\rangle, \label{gp}\\
G^-(x^\mu,y^\mu) = \langle \phi(y^\mu)  e^{-ie\int_x^{y} A_\mu dx'^\mu}   \phi^\dagger(x^\mu)\rangle, \label{gm}
\end{eqnarray}
where brackets indicate expectation value in the ``in" vacuum. On an equal time slice, we obtain
\begin{equation}
G^+(\eta;x,y) = G^-(\eta;x,y)=e^{-\lambda(y-x)/\eta} \int{dk\over 2\pi} |\phi_k^{\rm in}(\eta)|^2 e^{ik(y-x)}. \label{gpgm}
\end{equation}
From (\ref{asin}), we find that for fixed $\eta$ and large $|k|$, 
\begin{equation}
|\phi_k^{\rm in}(\eta)|^2 \sim {1\over 2|k|} \left[1+ O\left((k\eta)^{-1}\right)\right].
\end{equation}
The leading term is the same as for the Bunch-Davies modes, and so the integral in (\ref{gpgm}) leads to the standard logarithmic divergence.
Here, we have done point splitting on an equal time slice, but it is easy to check that the conclusion is the same if we split the points in an arbitrary direction.

Next, let us consider the current. This is given by
\begin{equation}
J_\mu =- {ie\over 2} \left(\phi^\dagger D_\mu \phi - \phi (D_\mu \phi)^\dagger\right) + {\rm h.c.}, \label{cu}
\end{equation}
where $D_\mu=\partial_\mu-ieA_\mu$. Its expectation value can be computed as
\begin{equation}
\langle J_\mu\rangle = -{ie\over 2} \lim_{x^\nu\to y^\nu}\left({\partial\over \partial y^\mu}-{\partial\over \partial x^\mu}\right)G^{(1)}(y^\nu-x^\nu) 
,
\end{equation}
where
\begin{equation}
G^{(1)}(x^\mu)=G^+(x^\mu)+G^-(x^{\mu}).
\end{equation}
Using (\ref{gpgm}), we have
\begin{equation}
\langle J_1\rangle = -2ie \lim_{x\to y}  \int_{-\infty}^{\infty} {dk\over 2\pi} \left[ik-{\lambda\over \eta}\right]|\phi_k(\eta)|^2 e^{(ik-\lambda/\eta)(y-x)}.\label{jexp}
\end{equation}
Also, it can be checked by direct substitution of (\ref{fieldexp}) into (\ref{cu}) that the charge density vanishes $\langle J_0\rangle=0$.
Using the asymptotic expansion of the Whittaker functions for large argument, we have 
\begin{equation}
|\phi_k|^2 = {1\over 2|k|}|W_{\pm\lambda,\sigma}(2i|k|\eta)|^2 e^{\pm i\pi\lambda} \approx {1\over 2|k|} \left|1 -{m^2/H^2\mp \lambda \over 2i|k|\eta} + 
O\left((k\eta)^{-2}\right)\right|^2,
\end{equation}
where the upper and lower signs correspond to $k>0$ and $k<0$, respectively.
Substituting this into (\ref{jexp}), we find that 
there is a linear divergence in momentum which is independent of the mass $m$, and no logarithmic divergence.

The divergence can be renormalized by means of a Pauli-Villars (PV) subtraction, involving a 
field of large mass $M$, which we will send to infinity after momentum integration,
\begin{equation}
J(E)\equiv  {1\over a(\eta)}\langle J_1\rangle_{ren} = {2e\over a} \lim_{M\to \infty}\left( \lim_{\Lambda\to \infty} \int_{-\Lambda}^{\Lambda} {dk\over 2\pi} \left[k+{i\lambda\over \eta}\right]\left[|\phi_k(\eta)|^2-|\phi_{k,M}(\eta)|^2\right]\right).\label{jren}
\end{equation}
Here, $\phi_{k,M}(\eta)$ are positive frequency modes of the ``in'' vacuum for the field of mass $M$.
The momentum integral is finite for any value of $M$, and we choose the limits of integration to be symmetric around $k=0$ for later convenience. For a field of large mass, we can use the WKB form for the mode function 
$\phi_{k,M}$,
\begin{equation}
|\phi_{k,M}(\eta)|^2 \approx |\phi_{k,M}^{WKB}(\eta)|^2= {1\over 2 \sqrt{M^2a^2+(k+i\lambda/\eta)^2}}. \quad (M^2\gg H^2, eE)
\end{equation}
This approximation becomes exact in the limit $M\to \infty$, and we can safely 
substitute $|\phi_{k,M}|^2$ by $|\phi_{k,M}^{WKB}|^2$ in Eq. (\ref{jren}). Note also that the contribution of the PV field to the current is actually independent of $M$ when we use $WKB$ mode functions,
\begin{equation}
\int_{-\Lambda}^{\Lambda} {dk\over 2\pi} \left[k+{i\lambda\over \eta}\right]|\phi^{WKB}_{k,M}(\eta)|^2 = 
	{1\over 4\pi}\left. \sqrt{M^2 a^2 +(k+i\lambda/\eta)^2}\right|^\Lambda_{-\Lambda} = {i\lambda\over 2\pi\eta}=-{|\lambda|\over 2\pi\eta}.
\label{pvcont}
\end{equation}
Substituting (\ref{pvcont}) into (\ref{jren}) and using (\ref{modes0}-\ref{modes}), we can rewrite the renormalized current as
\begin{equation}
J(E)= {eH\over \pi}\left(-{|\lambda|}+ \int_{0}^{+\infty} {dx\over 2x}\sum_{\pm} \left(|\lambda| \pm x\right) e^{\mp\pi |\lambda|}
\left|W_{\pm \lambda,\sigma}(-2ix)\right|^2\right) .\label{jnum}
\end{equation}
The first term is the contribution of the PV fields, and in the second term we have introduced $x=|k\eta|$ as the variable of integration.
The second term is actually finite if we perform the sum over positive and negative $k$ (i.e. the sum over $\pm$) before doing the integral, and so we can safely remove the cut-off $\Lambda$. 
With some ingenuity, the integral on the right hand side of Eq. (\ref{jnum}) can be computed analytically. This is done in Appendix \ref{integral}, where we show that
\begin{equation}
J(E) = {e\over \pi} {H\sigma\over \sin(2\pi\sigma)} \sinh(2\pi |\lambda|). \label{formula}
\end{equation}
Surprisingly, this agrees exactly with the semiclassical expression which we derived in Section \ref{analytic}, as can be seen by using Eq. (\ref{bogo}) for the Bogoliubov coefficients into Eq. (\ref{janalytic}).

\begin{figure}
\begin{center}
\vspace{-2cm}
\includegraphics[width=14cm]{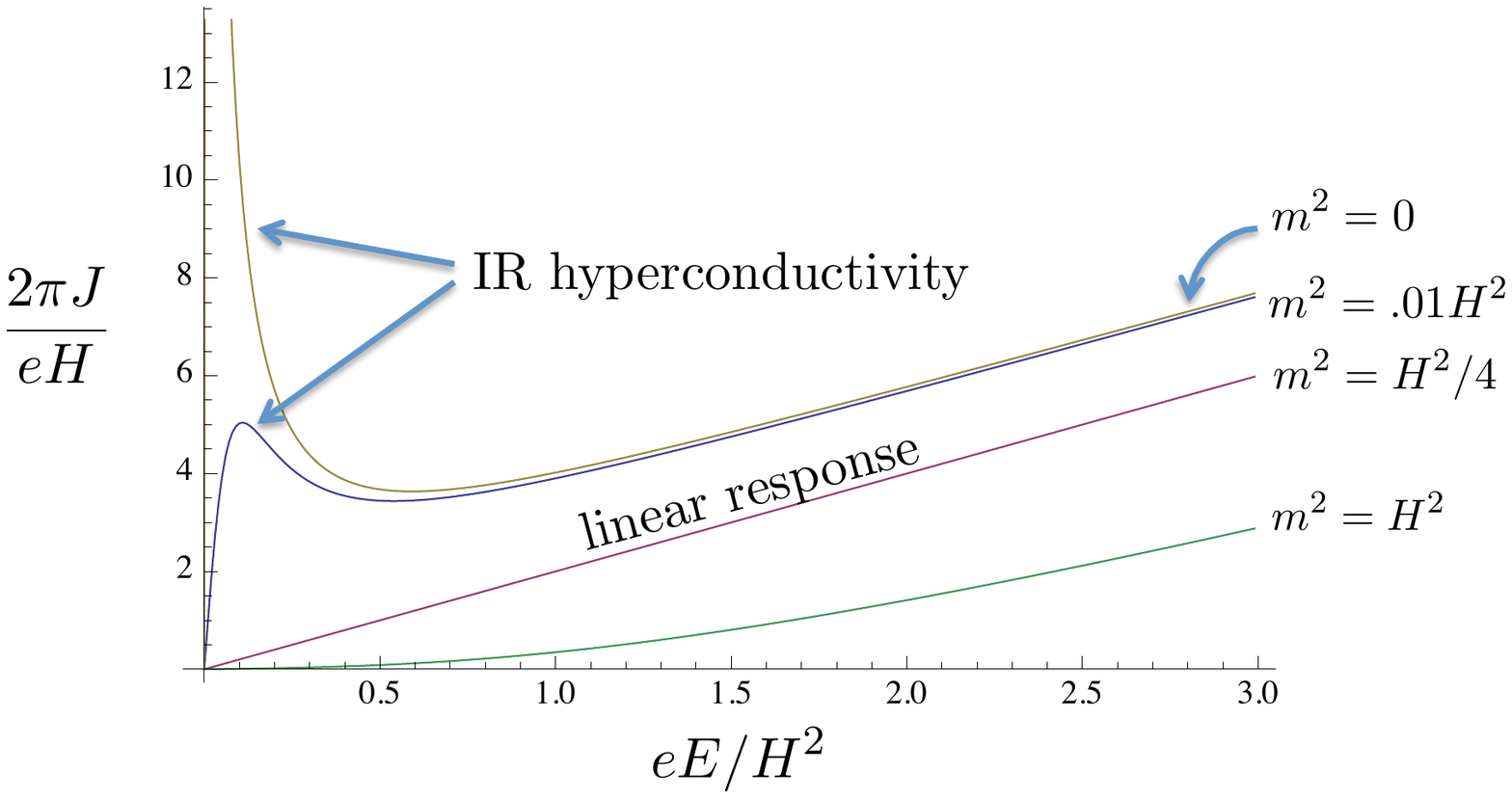}
\vspace{-1.5cm}
\caption{The renormalized current $J$ as a function of the electric field $E$, for different values of the mass $m$ of the charge carriers.}
 \label{current}
\end{center}
\end{figure}

In Fig. \ref{current} we plot the value of the current $J$ as a function of the electric field $E$, for different values of the mass. 
Let us now comment on the qualitative features of the current in different mass ranges.

\subsection{Linear response ($m^2=H^2/4$):} 

It follows from (\ref{formula}) that for $m^2 = H^2/4$ the current is exactly linear in the electric field:
\begin{equation}
J={e^2E\over \pi H}. \quad(m^2=H^2/4)\label{sc}
\end{equation}
Such linear response is reminiscent of the behaviour of currents due to massless charge carriers in flat space. 

The Schwinger pair creation rate for massless carriers in flat space is given by 
\begin{equation}
\Gamma={eE\over 2\pi},\label{frat}
\end{equation}
leading to a current which grows in time, at a constant rate which is proportional to the electric field $J=e^2E t/\pi$. In dS, we expect the current to be diluted by the expansion of the universe, and so the linear growth in time will be cut off. Naively replacing $t$ with the expansion time $H^{-1}$ leads to (\ref{sc}). 
More precisely, we may observe that the number density $n$ of charged pairs satisfies
\begin{equation}
{dn \over dt} = \Gamma - H n. \label{rateq}
\end{equation}
where the second term in (\ref{rateq}) accounts for cosmic dilution. This leads to the stationary solution
\begin{equation}
n={\Gamma\over H}.
\end{equation}
For massless (or highly relativistic) carriers, the current is 
\begin{equation}
J=2en=2e{\Gamma\over H}, \label{jmassless} 
\end{equation}
which agrees with (\ref{sc}) provided that we use the pair production rate (\ref{frat}).

The exact linearity in $eE$ seems nonetheless somewhat coincidental, particularly since in 1+1 dimensions the conformal value of the mass is $m^2=0$, while Eq. (\ref{sc}) holds for $m^2=H^2/4$. The latter value of the mass corresponds to the boundary where long wavelength modes behave as critically damped oscillators. For smaller values of the mass, infrared contributions to the current become important, as we shall now explain.

\subsection{IR hyperconductivity ($m^2\ll H^2$):}

A striking property of the regime $m^2\ll H^2$ is that for $eE\ll H^2$ the current is dominated by infrared contributions, rather than newly created pairs. 
This leads to a current of the form
\begin{equation}
J  \approx {1\over 2 \pi} {e^2 E H^3 \over m^2 H^2 + e^2 E^2}. \quad \left(eE \ll H^2 \right) \label{jsmall}
\end{equation}
This behaviour is illustrated in Fig. \ref{current}, for $m=0$ and $m=0.1 H$. 
The local maximum is at $eE \sim mH$, with $J \sim H^2/m$, and the current grows unbounded for small electric field in the 
limit $m\to 0$. Actually, for $m/H\ll eE/H^2 \ll1$
the current is inversely proportional to the applied electric field, 
\begin{equation}
J\sim {H^3\over E}, \quad (m/H\ll eE/H^2 \ll1)
\end{equation}
much in contrast with Ohm's law. Note that the current is also independent of the electric charge in this limit.

To understand the origin of (\ref{jsmall}), we first note that for small $z$, the Whittaker function has the behaviour
\begin{equation}
W_{\lambda,\mu} \approx z^\epsilon [1+O(z)], \quad (\epsilon \ll1) \label{Wz}
\end{equation}
where  we have introduced
\begin{equation}
\epsilon \equiv \left({m^2\over H^2} +{e^2 E^2\over H^4}\right).
\end{equation}
Next, from (\ref{jnum}) we see that the infrared contribution to the current comes from the first term in round brackets inside the integrand, and can be expressed as
\begin{equation}
J_{IR} = {2eH|\lambda|}\ \langle \phi^2 \rangle \label{jir}
\end{equation}
where here we have introduced
\begin{equation}
\langle \phi^2 \rangle \equiv {1\over 2\pi} \int_0^\infty {dx\over 2x} 
 \sum_{\pm} e^{\mp\pi |\lambda|} \left|W_{\pm \lambda,\sigma}(-2ix)\right|^2.
 \end{equation}
 Using (\ref{Wz}), and ignoring numerical coefficients, we can estimate
 \begin{equation}
 \langle \phi^2 \rangle \sim {1\over \epsilon}, \quad (\epsilon \ll1)
 \end{equation} 
 which substituted into (\ref{jir}) leads to (\ref{jsmall}). 
 
 An alternative heuristic derivation of (\ref{jsmall}) is the following. From the wave equation in the long wavelength limit
 it is easy to show that the non-decaying mode behaves as $\phi \propto e^{-\epsilon Ht}$. This means that in the absence of pair creation, the number of pairs
 would slowly dilute as $n \propto \phi^2 \propto e^{-2\epsilon Ht}$. Including pair creation at the rate $\Gamma$ per unit time and volume, we get
 \begin{equation}
 {dn \over dt} = \Gamma - 2\epsilon H n,
 \end{equation}
 which has the stationary solution
$ n=\Gamma/(2 \epsilon H)$.
This leads to 
\begin{equation}
J=2 e n =  e{ \Gamma \over \epsilon H},
\end{equation}
which coincides with (\ref{jsmall}) if we use $\Gamma\approx eE/(2\pi)$, which is the pair production rate for massless charge carriers in flat space.

 Since the infrared contribution can be very large for small mass and electric field, we will refer to this peculiar behaviour as
 infrared hyperconductivity. In general, the conductivity, defined as the ratio $J/E$, is larger for $m^2<H^2/4$ 
 than it is for the case with $m^2=H^2/4$, for all values of $eE$. Only for $eE \gg H^2$ do we recover 
 the linear response $J\approx eE/(\pi H)$.
 
 \subsection{Heavy pairs ($m\gtrsim H$):}
 
 In general, the current is suppressed as we increase the mass. We can distinguish two cases, according to the value of $\ell$. 
 
 \subsubsection{Cosmological pair production ($\ell \ll 1\lesssim m/H$)}
 
In this regime, the semiclassical action is given by 
\begin{equation}
S_{\pm} \approx 2\pi\left({m\over H} \pm |\lambda| \right) \gg 1. \label{useit}
\end{equation}
Pair production is exponentially suppressed, and so is the renormalized current. For very small electric field, $|\lambda| \ll 1$, the current is given by
\begin{equation}
J \approx 4  \left({m\over H}\right) {e^2 E\over H} e^{-2\pi m/H}.\quad\quad (|\lambda| \ll 1) \label{jsign}
\end{equation}
The presence of a Boltzmann suppression factor at the Gibbons-Hawking temperature $T=H/2\pi$ may naively suggest that gravitational particle production creates a hot plasma of charged particles, which are then set in motion by  the electric field, leading to a current. However, this interpertation would be rather imprecise. We will come back to this issue in Section \ref{conclusions}.

 \subsubsection{Pair production by the electric field ($\ell \gg 1$)}
 
In this limit, upward tunneling is very suppressed with respect to downward tunneling.
For $1\ll \ell\ll m/H$ the classical action is large and the acceleration time is much smaller than the Hubble time. In this case, 
an expression of the form (\ref{jmassless}) should be valid, where now $\Gamma$ is the flat space pair creation rate for massive particles,
given in (\ref{srate}),
 \begin{equation}
J \approx {e^2 E\over \pi H} e^{- \pi{m^2\over eE}}. \label{sch}
\end{equation}
This is indeed in agreement with Eq. (\ref{janalytic}) in the same limit. 
When the electric field is sufficiently large, $l\gg m/H$, the semiclassical action for tunneling, $S_- =-\pi m^2/eE$, is small and pair production is unsuppressed. This is illustrated in the bottom curve in Fig. \ref{current}, which shows that the current responds linearly to the electric field in this regime. In this sense, Eq. (\ref{sch}) can be extrapolated to very large electric field.

\section{Hadamard vacua and dS invariance}\label{hadamard}

We saw in Section \ref{currentsec} that the ``in" vacuum in the flat chart breaks dS invariance. We may ask whether this is due to a bad choice of the quantum state, 
or whether this feature is general and should be expected on physical grounds. After all, pair production induces the growth of a current. In this Section, we shall make this intuitive expectation more rigorous by showing that in any Hadamard vacuum dS invariance is broken.

It will be useful to think of 1+1 dimensional dS space as the hypersurface 
\begin{equation}
\eta_{AB} X^AX^B = H^{-2}, \quad    (A,B=0,1,2), \label{embedding}
\end{equation}
embedded in 2+1 dimensional Minkowski space with metric $\eta_{AB}={\rm diag}(-1,1,1)$. If $X^A$ and $Y^A$ are the coordinates of two points on this hypersurface, the variable
\begin{equation}
Z\equiv \cos\zeta \equiv H^2 X^AY_A,
\end{equation}
is dS invariant. If $X^A$ and $Y^A$ are spacelike separated, then $\zeta =Hd$ is real, and $d$ is the geodesic distance between the two points in dS.
If they are time-like separated, then $\zeta$ is purely imaginary and $|d|$ is the proper time separation along the geodesic connecting the two points. 
The dS metric can be written as
\begin{equation}
ds^2=H^{-2}d\zeta^2-\sin^2 \zeta d\tau^2 = H^{-2}{dZ^2\over (1-Z^2)} -(1-Z^2)d\tau^2,
\end{equation}
where $\zeta=0$ corresponds to some arbitrarily chosen base point $X^A$.

The electric field can be written in terms of the gauge potential
\begin{equation}
A_\mu = \sqrt{-g} \epsilon_{\mu\nu} \partial^{\nu}\sigma. \label{covgauge}
\end{equation}
as
\begin{equation}
E=\Box \sigma, \label{dalem}
\end{equation}
where $\Box$ stands for the covariant d'Alembertian and we use the convention $\epsilon_{\tau Z}=1$ for the Levi-Civita symbol. For a constant electric field, we can choose $\sigma=\sigma(Z)$, with
\begin{equation}
\partial_Z\left((1-Z^2)\partial_Z\sigma\right) = {E\over H^2}.
\end{equation}
Up to an irrelevant additive constant, the general solution of this equation is
\begin{equation}
\sigma = -{E\over H^2}\ln(1+Z) +C \ln\left({1+Z\over 1-Z}\right). \label{intconst}
\end{equation}
In order to have a regular gauge potential in the coincidence limit, $Z=1$, we choose $C=0$.

Let us now consider the two point functions $G^{\pm}$, defined in (\ref{gp}) and (\ref{gm}). Note that a Wilson line is inserted between the two points in order to make $G^{\pm}$ gauge invariant. If this is calculated along the geodesic which links the points $x$ and $y$, this specification of the path is dS invariant. Now, by using the covariant gauge
(\ref{covgauge}) with $\sigma=\sigma(Z)$, it is clear that the Wilson line vanishes
\begin{equation}
\int_x^y A_\mu dx^\mu =0. \label{vanishwl}
\end{equation}
The reason is that $A_Z=0$, while along a geodesic $dx^\mu=\delta^\mu_Z\ dZ$ . On the other hand, it is important to note 
that at the base point $x$ (corresponding to $Z=1$), the value of $\tau$ is completely
undefined, while $A_\tau =-EH^{-1}(1-Z) +2HC$ will only vanish at $Z=1$ provided that we choose $C=0$. In other words, the Wilson line is only well defined for this choice of the integration constant in (\ref{intconst}).

Using (\ref{vanishwl}), we see that in the covariant gauge, and with the dS invariant
specification of the path, $G^{\pm}$ coincides with the Wightman function.
This satisfies the standard wave equation for a charged field:
\begin{equation}
\left[(\nabla_\mu - ie A_\mu)(\nabla^\mu - ie A^\mu)-m^2\right]G^{\pm}(x,y) =0, \label{weq}
\end{equation}
where derivatives are with respect to the second argument, $y$.
Let us now look for  a dS invariant solution to (\ref{weq}), of the form
\begin{equation}
G(x,y) = G(Z),
\end{equation}
where from now on we drop the $\pm$ superscripts.
 Noting that $\nabla_\mu A^\mu=0$, $A_\mu\partial^\mu G(Z)=0$ and
 \begin{equation} 
 A_\mu A^\mu = -{E^2\over H^2}\left({1-Z\over 1+Z}\right),
  \end{equation}
 Eq, (\ref{weq}) reduces to
 \begin{equation}
(Z^2-1) {d^2G\over dZ^2}  +2 Z {dG\over dZ} +\left[{m^2\over H^2}+{e^2E^2\over H^4} \left({Z-1 \over Z+1}\right)\right] G=0.
\end{equation}
It should be noted that this equation is gauge invariant.\footnote{Here, we have derived it by using a specific form of the gauge potential
[i.e., Eq. (\ref{intconst}) with $C=0$] which is singular at $Z=-1$. However, it can be shown that the same equation is obtained by using a gauge potential which is everywhere regular.}
To determine the behaviour of $G$ in the coincidence limit, we look for solutions in a power series in the vicinity of $Z=1$,
\begin{equation}
G=(Z-1)^\alpha \sum_{n=0}^{\infty} a_n (Z-1)^n.
\end{equation}
The indicial equation $\alpha^2=0$ has a double root, and so there is a regular solution and a logarithmically divergent solution.
This is the expected behaviour for a two dimensional Green's function. 

However, we may also look at the behaviour of the solutions when the point $y$ is close to the antipodal point of $x$, corresponding to $Z=-1$. These can be expanded as
\begin{equation}
G=(Z+1)^\beta \sum_{n=0}^{\infty} b_n (Z+1)^n.
\end{equation}
In this case the indicial equation gives $\beta= \pm i eE/H^2$, and therefore the two point function necessarily has a branch cut singularity. This ``infrared" singularity is reminiscent of the case of a massless neutral field in dS, where the solutions of the second order equation for a dS invariant two point function are also singular at the antipodal point.\footnote{Mathematically, the situation for the case of a neutral massless field is somewhat different from the one we have here. For a neutral massless field, we have one solution which is regular both at $Z=1$ and at $Z=-1$, while the other one is singular at both points. For charged fields, both solutions are singular at $Z=-1$.} 
 In that case, it is known \cite{allen} that there is no dS invariant Fock vacuum, and we expect a similar situation in the present case.  Since a dS invariant two point function necessarily includes singularities of a type which is different from the Hadamard form, we conclude that there are no dS invariant Hadamard vacua for charged particles in the presence of an electric field.

\section{Persistence of memory}\label{persistence}

The current which we have obtained in Section \ref{current} selects a preferred time direction,
\begin{equation}
t^\mu \propto \epsilon^{\mu\nu} \langle J_\nu  \rangle,
\end{equation}
which is orthogonal to the frame in which the charge density vanishes $\langle J_0 \rangle =0$. 
An observer which is boosted with respect to $t^\mu$ will observe a non-vanishing charge density
$\langle J_{0'} \rangle \neq 0$.  Since the proper magnitude of the current tends to a constant, any effect of the preferred time direction will persist undiminished arbitrarily far into the future. 

Of course, this also happens in the case of flat space, where the current has the form (\ref{flatcurrent}). But, while in flat space we are used to the fact that initial conditions can have a lasting effect,
this may seem more surprising in an inflationary context. It is well known that a long period of inflation erases certain features of the initial conditions. For instance, 
unwanted relics are exponentially diluted away, and cosmological perturbations in the initial hypersurface (of unknown but possibly sizable amplitude) are stretched away to unobservably large distances.
While this is true, there are certain observables for which the influence of the initial hypersurface persists after an arbitrarily large period of inflation \cite{GGV}, and the current which we have discussed in this paper belongs to this category. The current is made out of positively charged particles accelerating towards the right, and negatively charged particles accelerating towards the left. If we are in the rest frame of initial conditions, the number of particles or antiparticles which will hit us from the left or from the right is the same. However, if we move towards the right, we are more likely to be hit by a charged particle which is coming from that direction.

The discussion of Ref. \cite{GGV} considered a simplified model of bubble nucleation, where the size of the bubbles at the time of nucleation was taken to be infinitessimally small. Here, we shall discuss a 
finite size effect, which has to do with the persistent influence of initial conditions in determining the frame of bubble nucleation\footnote{In the limit when the bubbles 
are point-like at the time of nucleation, there is no particular frame associated to the nucleation event.}. Before moving into the case of de Sitter, let us first briefly recall the situation in flat space. 

\subsection{Flat space}

It was found in Refs. \cite{GKSSV,GKT} that, in the Lorentz invariant ``in" vacuum, the frame of nucleation is very strongly correlated with the state of motion of the detector. Semiclassically, the trajectory of the two charges in a pair is given by the two branches of a hyperbola
\begin{equation}
x^2-t^2=r_0^2,    \label{trajectory}
\end{equation}
 where
\begin{equation}
r_0 = {m\over eE}.\label{rzero}
\end{equation}
The trajectory (\ref{trajectory}) has contracting and expanding phases, before and after the turning 
point at $t= 0$. In the frame of nucleation, which we may denote by $\tilde S$, the trajectory of the charged particles has the same form $\tilde x^2-\tilde t^2=r_0^2$
but only the expanding phase $\tilde t>0$ is physical: the particle and antiparticle nucleate at rest at $\tilde t=0$, and subsequently accelerate away from each other. 
In the frame of a detector consisting of a single particle, and moving at some speed $v$ relative to the frame of nucleation, the trajectory of the charged particles would again have the form $x'^2-t'^2=r_0^2$, 
but the physical half of it (with $\tilde t>0$) would now correspond to $t' > -v x'$. For $v\neq 0$, some of the contracting phase, with $t'<0$, would be visible to the detector. 
What was found, however, is that the detector only sees the expanding phase, with $t'>0$, and therefore both frames must coincide to very good accuracy. Quantitatively, the relative speed between the detector and the frame of nucleation was found to be bounded by \cite{GKT}
\begin{equation}
\label{relspeed}
\Delta v \sim S^{-1/3} \ll1.
\end{equation}
Here, $S = {\pi m^2/eE} \gg 1$ is the action of the instanton which describes pair creation.
The correlation between both frames is therefore very strong,\footnote{The bound (\ref{relspeed}) coincides with the minimum quantum uncertainty in the velocity of a non-relativistic charged particle embedded in
a constant electric field.  A velocity of order $\Delta v$ is reached after a time interval of order
$\Delta t \sim S_E^{-1/3} r_0 \ll r_0$ past the turning point.
If the interaction of the nucleated pair with the detector takes place in the vicinity of the turning point, the semiclassical description does not apply. But even in this case, it was found \cite{GKT} that there is a strong asymmetry in the momentum transferred from the nucleated particles to the detector, in the direction of expansion after the turning point, consistent with the detector seeing only the pairs moving away from each other .} at least in the case where the system is in the Lorentz invariant  (LI) ``in'' vacuum.

 ÊÊÊÊÊ 
Suppose now the false vacuum is prepared at time $t=0$, say, by turning on a 
constant electric field. This determines a preferred frame, $S$, which we call the frame of initial conditions, and so the system is no longer Lorentz invariant. 
After some transient behaviour, pairs will be produced at the Schwinger rate, for times \cite{GKT}
\begin{equation}
t\gg \tau_{nuc} \sim r_0. \label{bigt}
\end{equation}
Here, $\tau_{nuc}$ is the time it takes for a given pair to be excited out of the vacuum. 
This can be estimated to be of the same 
order as the size of the instanton, $r_0$, given in (\ref{rzero}).

Let $S'$ be the frame of a detector, moving at speed $v_d$ relative to $S$. In the new frame, the false vacuum region $t>0$ corresponds to
\begin{equation}
t' > -v_d x'.\label{tprime}
\end{equation}
For definiteness, let us choose $v_d>0$, 
with the detector following the world line $x'=0$.
The particle and antiparticle in a pair that nucleates at rest with respect to $S'$ will be initially at the locations 
\begin{equation}
x'_\pm = x'_0\pm r_0,
\end{equation}
where $x'_0$ is the midpoint between the two charges.
If the detector interacts with, say,  the positively charged particle at
$x=x'_+=0$ shortly after the time of nucleation, then the location of the negatively charged particle is at $x' \approx -2 r_0$. According to (\ref{tprime}), for the negatively charged particle 
to be in the false vacuum, we must have
\begin{equation}
\tau \gtrsim 2 v_d r_0. \label{bigtau}
\end{equation}
Here,
\begin{equation}
\tau=t' = \gamma_d^{-1}t,
\end{equation}
is the amount of proper time which the detector has spent in the false vacuum, with $\gamma_d=(1-v_d^2)^{-1/2}$ .

It follows from (\ref{bigt}) and (\ref{bigtau}) that if the detector has spent a short proper time $\tau$ in the false vacuum,
\begin{equation}
\tau \ll r_0, \label{smalltau}
\end{equation}
then this detector will feel the influence of initial conditions. Indeed, if the detector is non-relativistic, so that $\tau\sim t$, then (\ref{bigt}) is violated and there is not enough time for the electric field to produce a pair out of a vacuum fluctuation. On the other hand, if the detector is relativistic, there may be enough time, $t\gg r_0$, but then Eq. (\ref{smalltau}) is incompatible with (\ref{bigtau}), which tells us that the pairs will {\em not} be seen to nucleate in the rest frame of the detector. In both situations, the frame of initial conditions will have an appreciable effect.

Conversely, if the detector spends a large proper time
\begin{equation}
\tau \gg r_0 \label{largetau}
\end{equation}
in the constant electric field $E$, much larger than the size of the instanton, then we do not expect the initial hypersurface to play much of a role
in determining the frame of nucleation \cite{GKT}.  The condition  (\ref{largetau}) is trivially satisfied at sufficiently late times, for any given velocity $v_d$ of the detector, so we do not expect any influence of the initial conditions to survive in the asymptotic future. This is in agreement with the results which are obtained by using the Lorentz invariant ``in" 
vacuum. In that case, the electric field is switched on at past infinity, and the frame of nucleation is entirely determined by the state of motion of the detector \cite{GKSSV,GKT}.

\subsection{de Sitter}

Let us now consider the case of de Sitter. For simplicity, we focus on the case where the initial conditions are imposed on an equal time slice in the flat chart, $\eta=\eta_0$. 
A case of particular interest is the ``in" vacuum, which we can think of as a limiting case where $\eta_0\to -\infty$.

The embedding coordinates introduced in Eq.~(\ref{embedding}) are related to
the flat chart coordinates $(\eta, x)$ by
\begin{eqnarray}
U\equiv X^0-X^2 &=& {\eta^2-x^2\over \eta},\\
V\equiv X^0+X^2&=& -{1\over H^2 \eta},\\
X^1&=& -{x\over H\eta}.
\end{eqnarray} 
The trajectory of a charged pair can be obtained by 
intersecting the hyperboloid (\ref{embedding}) with the plane \cite{j2}
\begin{equation}
X^2=w_0^\pm. \label{planewpm}
\end{equation}
Here 
\begin{equation}
w_0^{\pm} =\mp (H^{-2}-R^2_0)^{1/2},  \label{wopm}
\end{equation}
with
\begin{equation}
R_0^2 = {m^2\over m^2H^2 + e^2 E^2}.
\end{equation}
In (\ref{planewpm}), the plus sign corresponds to upward tunneling, and the minus sign to downward tunneling.
The intersection of (\ref{embedding}) and (\ref{planewpm}) leads to hyperbolas in the $(X^0,X^1)$ plane, of the form
\begin{equation}
(X^1)^2-(X^0)^2 = R_0^2. \label{hyperbola}
\end{equation}
The two branches of (\ref{hyperbola}) correspond to the worldlines of the two charges in the pair.
In terms of the flat chart coordinates, these worldlines are given by
\begin{equation}
x^2=H^{-2}(1+H^2\eta^2)+2w_0^{\pm}\eta. \label{standardf}
\end{equation}
The center of symmetry of the trajectory is at the point $X^0=X^1=0$, and $X^2=H^{-1}$. 
In the flat chart coordinates, this corresponds to the spacetime point
\begin{equation}
\eta=-H^{-1}, \quad x= 0. \label{standard} 
\end{equation}
We may refer to this center of symmetry as the ``nucleation event", although
strictly speaking nucleation takes up an extended region of spacetime. 
By using SO(2,1) transformations, the trajectory of any nucleated pair can be brought to the ``standard" form 
(\ref{standardf}), where the center of symmetry is at (\ref{standard}), so without loss of generality we shall restrict attention
to this semiclassical trajectory.

The physical momentum of each one of the particles in the pair is given by
\begin{equation}
p^{\pm}={mx'\over \sqrt{1-x'^2}}=\pm {eE\over H} \left(1+{m^2 H^2\over e^2 E^2}\right)^{1/2}{1\over a}+ {eE\over H}, \label{psem}
\end{equation}
where the superindex $\pm$ refers to the solution for upward or downward tunneling, respectively. Comparing this semiclassical expression
for the momentum with Eq. (\ref{pk}) we find that the trajectories (\ref{standardf}) correspond to modes with
\begin{equation}
k\approx \pm k_\sigma\equiv \pm H|\sigma|,
\end{equation}
where we have used $|\sigma|\gg 1$. According to (\ref{tk}), the time at which these modes are excited corresponds to $a(\eta_{k_\sigma})=1$.
This suggests that the semiclassical trajectory (\ref{standardf}) should be restricted to 
\begin{equation}
\eta>\eta_{k_\sigma}=-H^{-1}.Ê
\end{equation}
This would correspond to a pair nucleating on an equal time hypersurface, with the particle and antiparticle materializing at the same value of $\eta$.

For the flat chart ``in" vacuum state, which breaks dS invariance, the $\eta=const.$ hypersurfaces 
correspond to the preferred frame which is determined by the initial conditions. As mentioned at the beginning of this Section, 
it is reasonable to expect that such initial conditions may have some influence in determining the rest frame of nucleation. 
We saw that, in flat space, initial conditions may have some impact, but this fades away in 
the asymptotic future. Let us now show that things can be quite different in dS space.

First, we note that for an inertial detector which is not co-moving, the proper time which has been spent in the flat chart of dS is bounded by \cite{BGV}
\begin{equation}
\tau \leq {1\over 2H} \ln\left({\gamma_d+1\over \gamma_d-1}\right).\label{bgv}
\end{equation}
Here, $\gamma_d$ is the relativistic factor of the detector relative to the co-moving observers.
Consider now the situation where $m^2H^2 \ll e^2E^2$. In this case 
\begin{equation}
R_0\approx {m\over eE} =r_0
\end{equation}
is approximately equal to the flat space value $r_0$ for the radius of the hyperbola, with
\begin{equation}
r_0 \ll H^{-1}.
\end{equation}
It follows that for a highly relativistic detector, with 
\begin{equation}
\gamma_d\gg {1\over Hr_0} \label{expectat}
\end{equation}
we have 
\begin{equation}
\tau\ll r_0. \label{smalltime}
\end{equation}
As we saw in the case of flat space, a detector which has been in the false vacuum for such a short amount of time can feel the influence of initial conditions.
Pairs are not necessarily expected to nucleate in its rest frame.

A second point to note is that, in dS, we should be specific about what we mean by pairs nucleating at rest in a given frame. For a pair nucleating on the hypersurface $\eta=-H^{-1}$, the physical 
momentum of the particles at the time of nucleation, relative to the co-moving observers, is given by (\ref{psem}) with $a=1$. This is non-vanishing, both for upward and for downward tunneling. 
Because of that, instead of asking whether the pair nucleates at rest in the frame of the detector, it may be more pertinent to ask whether a boost 
in the detector's worldline is accompanied by a corresponding boost in the hypersurface on which the two charges in the pair are seen to nucleate simultaneously.

For a pair nucleating on the $\eta=-H^{-1}$ hypersurface, the temporal coordinate of the particles on the hyperbola (\ref{hyperbola}) at the time of nucleation is given by
\begin{equation}
X_{initial}^0=H^{-1}-w_0^\pm >0.
\end{equation}
Boosts in the $X^1$ direction will change initial value $X_{initial}^0$, without changing the form of the semiclassical trajectory (\ref{hyperbola}).
Introducing the boost parameter $\phi_1$ through the relation 
\begin{equation}
X_{initial}^0 = R_0 \sinh \phi_1,
\end{equation}
a boost of velocity $v=\tanh\phi_1$ in the $(X^0,X^1)$ plane will bring the initial time in the trajectory of one of the charges to the value $X^0=0$. By further 
increasing the boost parameter, this initial time will go into negative values of $X^0$. However, in the flat chart, there is a minimum value of $X^0$ on the trajectory 
of the pair, given by 
\begin{equation}
X^0>X^0_{min}\equiv -w_0^{\pm}.
\end{equation}
Lower values of $X^0$ are outside of the flat chart. The boost parameter $\phi_2$ which is needed to bring the initial time of the particle trajectory from $X^0=0$ to 
$X^0=-w_0^{\pm}$ is given by
\begin{equation}
w_0^\pm = R_0 \sinh\phi_2.
\end{equation}
The maximum boost which can be applied to the pair which nucleates on the $\eta=H^{-1}$ hypersurface without having one of the particles in the pair 
start its worldline outside of the flat chart is given by
\begin{equation}
\phi_{max}=\phi_1+\phi_2.
\end{equation}
Note that
\begin{equation}
\sinh\phi_1+\sinh\phi_2 ={1\over HR_0}.
\end{equation}
For $H R_0\sim 1$, we have
\begin{equation}
\gamma_{max} \equiv \cosh(\phi_1+\phi_2) \sim 1.
\end{equation}
For $H R_0 \ll 1$, 
\begin{equation}
\gamma_{max} \approx {1\over H R_0}
\end{equation}
for downward tunneling, and $\gamma_{max} \sim 1$ for the case of upward tunneling.

Ignoring upward tunneling, we conclude that a detector with relativistic factor 
\begin{equation}
\gamma_d\gg\gamma_{max} \sim {1\over H R_0} \label{gammamax}
\end{equation}
with respect to the co-moving frame, cannot detect pairs whose nucleation hypersurface is
boosted by the same relativistic factor relative to the $\eta=const.$ hypersurface. Including upward tunneling, the same conclusion applies
for $\gamma_d\gg 1$. In this sense, fast moving detectors feel the influence of the hypersurface of initial 
conditions in which the false vacuum has been prepared. Note that this conclusion is in agreement with our earlier expectation, which was based on 
Eqs. (\ref{expectat}) and (\ref{smalltime}). 

Unlike the case of flat space, here the influence of initial conditions in determining the frame of nucleation persists arbitrarily far into the future. Note that
here we have considered pairs nucleating at $\eta=-H^{-1}$, but since the surface of initial conditions is $\eta\to -\infty$, our analysis, and the estimate of 
$\gamma_{max}$ given in (\ref{gammamax}) is independent of the time at which nucleation occurs. 


\section{Summary and discussion}\label{conclusions}

Vacuum transitions in an inflating multiverse may proceed by quantum tunneling. A simple model where such transitions can be analyzed beyond the semiclassical approximation is the Schwinger process
in 1+1 dimensions. 

In de Sitter space, both the electric and gravitational fields can pull pairs out of the vacuum. Particles and antiparticles are subsequently accelerated by the electric field and diluted by cosmic 
expansion. This results in a stationary spacelike electric current of proper magnitude $J$,  given by Eq. (\ref{formula}), as
\begin{equation}
J_\phi= {1\over \pi} {H\sigma\over \sin(2\pi\sigma)} \sinh(2\pi e E/ H^2). \label{f1}
\end{equation}
Here $J_\phi=J/e$ is the charge number current, and
\begin{equation}
\sigma=\left({1\over 4}-{m^2\over H^{2}}-{e^2E^2\over H^{4}}\right)^{1/2}.
\end{equation}
Throughout this paper the electric and gravitational fields have been treated as external sources. In the situation where the electric field is dynamical, 
Gauss's law requires a discontinuity $\Delta E=e$ accross the position of the charges.   
The decoupling limit where the electric field can be treated as external corresponds to $e/E\to 0$ while keeping $eE$ fixed. \footnote{There is no Einstein gravity in $1+1$ dimensions and in this case there is no need to decouple it. In a more general context, gravity can also be decoupled by taking the limit where Newton's constant vanishes while keeping the dS radius $H^{-1}$ finite. }
Within this limit, Eq. (\ref{f1}) is valid in the full range of parameters $m$, $eE$, and $H\neq 0$. 
\footnote{We note, in passing, that the definition of a pair production rate per unit volume, $\Gamma$, is somewhat ambiguous in dS. The reason is that the volume grows with time in an expanding universe, while the time of pair creation is not defined very precisely. By contrast, the expectation value of the current $J_\phi$ does not suffer from this ambiguity, and can therefore be thought of as a more precise characterization of the Schwinger process.} 

A non-vanishing current breaks dS invariance. Since the background is invariant, this has to be attributed to the choice of quantum state. 
In our case, the current is parallel to the equal time slicing in the flat chart, which is used in order to define the ``in" vacuum.\footnote{Pair production in the ``in" vacuum of the global chart of dS, which includes the 
contracting and expanding phases, has been considered in Ref. \cite{KP}. Even in the absence of an electric field, such ``in" vacuum is not Hadamard (see e.g. Ref. \cite{BMS}). We expect 
that the presence of the electric field will only make things worse, since pairs can be produced for an infinite amount of time in the contracting phase of de Sitter, resulting in an infinite density.}
More generally, we have shown that it is not possible to choose a dS invariant Hadamard 
quantum state for $E\neq 0$.

The semiclassical regime corresponds to imaginary values of $\sigma$, with $|\sigma\pm\lambda|\gtrsim 1$, where $\lambda=i eE/H^2$. In this case, the 
 distribution of created particles is given by \cite{j},
\begin{equation}
{dN\over dx} = |\beta^{\pm}|^2 {dk\over 2\pi}, \label{dsummary}
\end{equation}
where $\beta^\pm$ are the Bogoliubov coefficients, given in Eq. (\ref{bogo}), and  the double sign refers to positive and negative values of the co-moving momentum $k$, respectively.
These, in turn, correspond to upward and downward transitions. The distribution (\ref{dsummary})  is discontinuous at $k=0$, but it is otherwise flat, with a UV cut-off which depends on time, 
\begin{equation}
|k| < k_c^\pm = a(t) \left(p_c^\pm\mp{eE\over H}\right). \label{cut}
\end{equation}
Here $a(t)$ is the scale factor, and $p^{\pm}_c$ is the physical momentum of the particles at the  time of pair creation\footnote{\label{adi} The time of pair creation is estimated as the time when the violation of adiabaticity in the evolution of the mode functions is maximum.} 
\begin{equation}
p_c^\pm \sim |\sigma| H \pm {eE\over H}.\label{pcs}
\end{equation}
It was shown in Refs. \cite{j,j2} that Eq. (\ref{dsummary}) is in agreement with the distribution of particles which can be calculated with instanton methods. 

Interestingly, the exact expression for the renormalized expectation value of the current, Eq. (\ref{f1}), can also be obtained from a simple semiclassical computation, where we add the contributions
from all individual pairs in the distribution (\ref{dsummary}). Aside from the current flowing along the semiclassical 
trajectories, a vacuum current has to be included, connecting the particle and antiparticle at the time of pair creation, so that charge is locally conserved.
It turns out that the contribution from the vacuum current is comparable to that from the semiclassical trajectories, 
and the sum of these two is insensitive to the precise value which we adopt for $p_c^{\pm}$  
[as long as we take $k_c \propto a(t)$, as in (\ref{cut})].
The total current takes then the form
\begin{equation}
J_\phi = {H |\sigma| \over \pi} \left(|\beta^-|^2 - |\beta^+|^2\right), \label{janalytic2}
\end{equation}
which manifestly shows the separate contributions from downward and upward transitions. This expression reproduces (\ref{f1}) exactly once we substitute the Bogoliubov coefficients given by (\ref{bogo}).

The conductivity of the vacuum in different regimes can be readily analyzed from (\ref{f1}). For $m^2 = H^2/4$, we find a linear response $J_\phi = eE/(\pi H)$, with resistivity proportional to the expansion rate $H$.
This particular value of the mass corresponds to the case where (in the absence of the electric field) long wavelength modes behave as critically damped oscillators. 

For smaller values of the mass, $m^2 \ll  H^2/4$, infrared effects are important and
the conductivity can be very large. In fact, for $m H \ll eE \ll H^2$, the current behaves in inverse proportion to the electric field, $J_\phi \approx H^3 / (2\pi eE)$, much in contrast with Ohm's law. This phenomenon is due to the infrared behaviour of the two point function in dS, and is expected to be present also in 3+1 dimensions, for light fields with $m \ll H$. Infrared hyperconductivity may have important consequences for cosmology (e.g. in scenarios where magnetic fields are generated during inflation). We leave this as a subject for future research.

For $m\gg H$, $\sigma$ is imaginary, with $|\sigma|\gg1$. The current can then be estimated from (\ref{janalytic2}),
with
\begin{equation}
|\beta^\pm|^2 \approx e^{-2\pi(|\sigma|\pm |\lambda|)}.
\end{equation}
Here $|\lambda|=eE/H^2$. For $|\lambda|\ll |\sigma|$, gravitational pair production is more important than production by the electric field. In particular, 
for very small electric field $2\pi |\lambda| \ll 1$, we have
\begin{equation}
J_\phi  \approx 4 {eE \over H}  {m\over H} e^{-{2\pi m\over H}}. \label{strange}
\end{equation}
The presence of a Boltzmann suppression factor at the Gibbons-Hawking temperature $T=H/2\pi$ may naively suggest that gravitational particle production creates a hot plasma of charged particles, which are then set in motion by  the electric field, leading to a current. However, this interpretation would be rather imprecise. The terminal velocity of charged particles in the expanding universe due to the applied electric field is of order $v_E \sim eE/(mH) \ll1$, while the number density of particles in a thermal bath at temperature $T$ is given by $n_T \sim (mT)^{1/2}  e^{-m/T}$. Eq. (\ref{strange}) is {\em not} of the form $J_\phi^{E,T} \sim v_E n_T$. Rather, the actual current is much larger, and follows from the flat distribution (\ref{dsummary}), which is not at all thermal. In particular, for large mass $m\gg H$ and small electric field $eE \ll m H$, the distribution contains highly relativistic particles up to the momentum cut-off (\ref{pcs})\footnote{As mentioned around Eq. (\ref{janalytic2}), the semiclassical estimate of the current is insensitive to the precise value of $p_c$. Ignoring, for the sake of argument, the adiabaticity criterion (see footnote \ref{adi}), one might think that  $p_c$ could be taken to be non-relativistic. However, as explained in Appendix \ref{app}, if the electric field is small, then by the time the momentum is non-relativistic the particle and antiparticle are already separated by a distance much larger than $H^{-1}$. Therefore, a non-relativistic value of  $p_c$ seems to contradict the notion that pair creation is a local process. By contrast, the estimate (\ref{pcs}) corresponds to the time when the distance between particle and antiparticle is near its minimum (and is at most of order $H^{-1}$).}, while the thermal distribution would be non-relativistic in this parameter range.

This is somewhat puzzling, since it is well known that in the limit $E\to 0$ a particle detector responds as if it were immersed in a thermal bath (see Appendix \ref{detector}). The response of a detector does not necessarily reflect the existence of actual particles\footnote{For instance, a co-moving detector in an expanding universe will be excited even in a conformal vacuum, where (in the absence of the detector) particle creation does not occur.  Also, an accelerated detector responds thermally even in a Minkowski vacuuum.}, but in our case it would be good to understand the reason why the flat distribution up to relativistic values of the momentum
is not detected.  Note also that, in the absence of an electric field, we can always 
choose a dS invariant state for $\phi$. However, the cut-off (\ref{cut}) in the semiclassical distribution (\ref{dsummary}) clearly breaks dS invariance regardless of the value of $E$. An explicit resolution of this puzzle, particularly in the limit $E\to 0$, would be very interesting and is left for further research.
Meanwhile, we note that in the absence of an electric field, the current vanishes, so there is no contradiction between dS invariance and the observable (\ref{strange}).

In general, the non-vanishing current selects a preferred time direction,
\begin{equation}
t^\mu \propto \epsilon^{\mu\nu} \langle J_\nu  \rangle,
\end{equation}
which is orthogonal to the frame in which the charge density vanishes $\langle J_0 \rangle =0$. An observer which is boosted with respect to $t^\mu$ will see a non-vanishing charge density
$\langle J_{0'} \rangle \neq 0$.  Since the proper magnitude of the current tends to a constant, any effect of the preferred time direction will persist undiminished arbitrarily far into the future. 
The current is made out of positively charged particles accelerating towards the right, and negatively charged particles accelerating towards the left. If we are in the rest frame of initial conditions, the number of particles or antiparticles which will hit us from the left or from the right is the same. However, if we move towards the right, we are more likely to be hit by a charged particle which is coming from that direction. 
This is the persistence of memory effect first discussed in \cite{GGV}. 
Here, we have argued that initial conditions will also have a persistent influence in determining the frame of nucleation of new pairs. The reason is simple. Denoting by $R_0$ the size of the instanton, the frame of nucleation cannot be boosted by a relativistic factor larger than $\gamma \gtrsim (HR_0)^{-1}$ relative to the frame of initial conditions, without having one of the particles in the pair intersect the hypersurface of initial conditions.

\acknowledgments

J.G., S.K. and A.V. are grateful to YITP for hospitality during the long term workshop YITP-T-12-03. 
This work was supported in part by grant AGAUR 2009-SGR-168, MEC FPA 2010-20807-C02-02, CPAN CSD2007-00042 Consolider-Ingenio 2010, PHY-1213888
from the National Science Foundation, the Grant-in-Aid for Scientific Research (Nos. 21244033, 21111006, 24103006, 24103001 and 25400251), the Grant-in-Aid for the Global COE Program "The Next Generation of Physics, Spun from Universality and Emergence" from the Ministry of Education, Culture, Sports, Science and Technology of Japan and funding from the University Research Council of the
University of Cape Town. M.B.F. is supported by a FPU scholarship no. AP2010-5453.

\appendix

\section{Semiclassical trajectory parametrized by momentum} \label{app}

The semiclassical trajectory of a charged pair in the flat chart of dS, Eq. (\ref{standardf}), can be parametrized in terms of the physical momentum $p$ with respect to the co-moving congruence, given in (\ref{psem}). From these two equations, it is straightforward to check that the physical distance between the two members of a pair is given by
\begin{equation}
d^2 \equiv (2 a\ x)^2 = 4 H^{-2} {m^2 + p^2 \over \left(p-{eE\over H}\right)^2}. 
\end{equation} 
Using
\begin{equation}
p={k\over a} +{eE \over H},
\end{equation}
we immediately obtain Eq. (\ref{distance}) in the main text.

For $k>0$, the physical distance between the charges stays approximately constant $d\approx 2 H^{-1}$ while the particles
are relativistic with respect to the co-moving congruence  $(k/a)\sim p \gg m$. The distance grows in proportion to the scale factor, after the momentum becomes non-relativistic. This is in agreement with the idea that pairs are created with $p \sim m$, since it would be hard to create them afterwards, when the particle and antiparticle are separated by many horizon regions. 
For $k<0$ the minimum distance is for $p=-m/l$, where $l\equiv eE/(mH)$, and it is given by
\begin{equation}
d= {2 H^{-1} \over (1+l^2)^{1/2}}.
\end{equation}
For small electric field, $l \ll1$, the distance stays of order $d\sim 2H^{-1}$ while the charges are relativistic, and increases exponentially afterwards. For large electric field $l \gg 1$, the distance is of order $d\sim 2m/eE\ll 2H^{-1}$ for $p\approx 0$.
Again, the time when the physical distance between the particle and antiparticle in a pair is near its minimum is in agreement with our earlier estimate (\ref{tk}) for the time of pair creation.

\section{Momentum integral}\label{integral}

Here, we calculate the integral in Eq. (\ref{jnum}), which gives the expectation value of the current in the ``in'' vacuum:
\begin{equation}
J = \frac{eH}{\pi} \left[ - \abs{\lambda} + \int_0^\infty \left( \sum_{\pm} \left( \abs{\lambda} \pm x \right) \mathe^{\mp \pi \abs{\lambda}} \abs{ W_{\pm \lambda, \sigma}(-2 i x) }^2 \right) \frac{\total x}{2x} \right] \eqend{.}
\end{equation}
with
\begin{equation}
\lambda = i \frac{e E}{H^2} \eqend{,} \qquad \sigma = \sqrt{\frac{1}{4} - \frac{m^2}{H^2} - \frac{e^2 E^2}{H^4}} \eqend{.}
\end{equation}
The integral is convergent, given that $\lambda = i \abs{\lambda}$ is purely imaginary, but for the manipulations in the following it is necessary to temporarily insert a factor $\mathe^{-\epsilon x}$ to make each term converge individually.

The Whittaker function can be written as
\begin{equation}
W_{\pm \lambda, \sigma}(-2 i x) = \mathe^{i x - i \frac{\pi}{4} \left( 1+2\sigma \right)} (2 x)^{\frac{1}{2}+\sigma} U\left( \frac{1}{2}+\sigma \mp \lambda, 1+2\sigma, -2 i x \right) \eqend{.}
\end{equation}
with the confluent hypergeometric function $U$. This function admits the integral representation
\begin{equation}
U(a,b,z) = \frac{1}{\Gamma(a)} \int_0^\infty \mathe^{-zt} t^{a-1} (1+t)^{b-a-1} \total t
\end{equation}
for $\Re a > 0$ and $\arg z < \frac{\pi}{2}$, from which we see that
\begin{equation}
\begin{split}
\abs{W_{\pm \lambda, \sigma}(-2 i x)}^2 &= 2 x (- 2 i x)^\sigma (2 i x)^{\sigma^*} \ \ \times \\
&U\left( \frac{1}{2}+\sigma \mp \lambda, 1+2\sigma, -2 i x \right) U\left( \frac{1}{2}+\sigma^* \pm \lambda, 1+2\sigma^*, 2 i x \right) \eqend{.}
\end{split}
\end{equation}
We thus obtain
\begin{equation}
J = - \frac{eH}{\pi} \abs{\lambda} + \frac{eH}{\pi} \lim_{\epsilon \to 0} K,
\end{equation}
with
\begin{equation}
\begin{split}
K &= \int_0^\infty (-2i x)^\sigma (2 i x)^{\sigma^*} \mathe^{-\epsilon x} \ \ \times\\
& \bigg[ \left( \abs{\lambda} + x \right) \mathe^{- \pi \abs{\lambda}} U\left( \frac{1}{2}+\sigma - \lambda, 1+2\sigma, -2 i x \right) U\left( \frac{1}{2}+\sigma^* + \lambda, 1+2\sigma^*, 2 i x \right) \\
&\qquad+ \left( \abs{\lambda} - x \right) \mathe^{\pi \abs{\lambda}} U\left( \frac{1}{2}+\sigma + \lambda, 1+2\sigma, -2 i x \right) U\left( \frac{1}{2}+\sigma^* - \lambda, 1+2\sigma^*, 2 i x \right) \bigg] \total x \eqend{.}
\end{split}
\end{equation}

We now use the Mellin-Barnes representation for the confluent hypergeometric function
\begin{equation}
U(a,b,z) = \int_\mathcal{C} \frac{\Gamma(a+s) \Gamma(-s) \Gamma(1-b-s)}{\Gamma(a) \Gamma(a-b+1)} z^s \frac{\total s}{2 \pi i}
\end{equation}
for $\abs{\arg z} < \frac{3}{2} \pi$. After shifting the integration variables $s \to s - \sigma$, $t \to t - \sigma^*$, this gives
\begin{equation}
\begin{split}
K = \int_0^\infty \mathe^{-\epsilon x} & \Bigg[ \left( \abs{\lambda} + x \right) \mathe^{- \pi \abs{\lambda}} \int_\mathcal{C} \frac{\Gamma\left( \frac{1}{2} - \lambda + s \right) \Gamma(\sigma-s) \Gamma(-\sigma-s)}{\Gamma\left( \frac{1}{2} - \lambda - \sigma \right) \Gamma\left( \frac{1}{2} - \lambda + \sigma \right)} (-2 i x)^s \frac{\total s}{2 \pi i} \\
&\qquad\times \int_\mathcal{C} \frac{\Gamma\left( \frac{1}{2} + \lambda+t \right) \Gamma(\sigma^*-t) \Gamma(-\sigma^*-t)}{\Gamma\left( \frac{1}{2} + \lambda - \sigma^* \right) \Gamma\left( \frac{1}{2} + \lambda + \sigma^* \right)} (2 i x)^t \frac{\total t}{2 \pi i} \\
&+ \left( \abs{\lambda} - x \right) \mathe^{\pi \abs{\lambda}} \int_\mathcal{C} \frac{\Gamma\left( \frac{1}{2} + \lambda+s \right) \Gamma(\sigma-s) \Gamma(-\sigma-s)}{\Gamma\left( \frac{1}{2} + \lambda - \sigma \right) \Gamma\left( \frac{1}{2} + \lambda + \sigma \right)} (-2 i x)^s \frac{\total s}{2 \pi i} \\
&\qquad\times \int_\mathcal{C} \frac{\Gamma\left( \frac{1}{2} - \lambda+t \right) \Gamma(\sigma^*-t) \Gamma(-\sigma^*-t)}{\Gamma\left( \frac{1}{2} - \lambda - \sigma^* \right) \Gamma\left( \frac{1}{2} - \lambda + \sigma^* \right)} (2 i x)^t \frac{\total t}{2 \pi i} \Bigg] \total x \eqend{.}
\end{split}
\end{equation}
The integration contours run from $- i \infty$ to $+ i \infty$, separating left from right poles. Since $0 \leq \Re \sigma < \frac{1}{2}$, this means we can choose both $s$ and $t$ contours to run parallel to the imaginary axis between $- \frac{1}{2} < \Re s, \Re t < - \Re \sigma$.

Because of the convergence factor, we can interchange the integrations. The integral over $x$ is elementary (since $\Re (s+t) > -1$ it converges at $x=0$). Taking into account that since $\sigma$ is either real or purely imaginary, we can replace all $\sigma^* = \pm \sigma$ by symmetry and obtain
\begin{equation}
\begin{split}
K &= \frac{\cos(2\pi\lambda) + \cos(2\pi\sigma)}{8 \pi^2} \int_{\Re s = - (1+2\Re\sigma)/4} \Gamma(\sigma-s) \Gamma(-\sigma-s) \int_{\Re t = - (1+2\Re\sigma)/4} \Gamma(\sigma-t) \Gamma(-\sigma-t)  \\
&\bigg[\Gamma\left( \frac{1}{2} - \lambda + s \right) \Gamma\left( \frac{1}{2} + \lambda+t \right) \mathe^{- i \frac{\pi}{2} (s-t-2\lambda)} (1+s+t- i \epsilon \lambda) \\
&\qquad- \Gamma\left( \frac{1}{2} + \lambda+s \right) \Gamma\left( \frac{1}{2} - \lambda+t \right) \mathe^{- i \frac{\pi}{2} (s-t+2\lambda)} (1+s+t+ i \epsilon \lambda) \bigg]\\
& \Gamma(1+s+t) \left( \frac{\epsilon}{2} \right)^{-2-s-t} \frac{\total t}{2 \pi i} \frac{\total s}{2 \pi i} \eqend{.}
\end{split}
\end{equation}

For the integral over $t$, we shift the contour over the poles at $t = -\frac{1}{2}\pm\lambda$, $t = -\frac{3}{2}\pm\lambda$, $t=-1-s$ and $t=-2-s$. The remaining contour integral is then bounded uniformly in $\epsilon$ and vanishes as $\epsilon \to 0$, so that the result is given by the residues of these poles and we obtain
\begin{equation}
\begin{split}
K &= \frac{\cos(2\pi\lambda) + \cos(2\pi\sigma)}{8 \pi^2} \int_{\Re s = - (1+2\Re\sigma)/4} \Gamma(\sigma-s) \Gamma(-\sigma-s) \sum_{\pm} \Gamma^2\left(\frac{1}{2}+s\pm\lambda\right) \bigg[ \\
&+ \mathe^{-i \pi (s\pm\lambda)} \left( (s+1)^2 - \sigma^2 \pm \lambda (3\pm2\lambda+2s) \right) \frac{\Gamma\left(-\frac{3}{2}-s\mp\lambda \right)}{\Gamma\left(\frac{1}{2}+s\pm\lambda \right)} \Gamma(1+s-\sigma) \Gamma(1+s+\sigma) \\
&\mp (1-i) 2^{s\pm\lambda} \mathe^{- i \frac{\pi}{2} (s\pm\lambda)} \epsilon^{-\frac{3}{2}-s\mp\lambda} (1+2s\pm2 \lambda) \Gamma\left(\frac{1}{2}\mp\lambda-\sigma\right) \Gamma\left(\frac{1}{2}\mp\lambda+\sigma\right) \\
&\mp (1+i) 2^{-2+s\pm\lambda} \mathe^{- i \frac{\pi}{2} (s\pm\lambda)} \epsilon^{-\frac{1}{2}-s\mp\lambda} \left( (1\pm2 \lambda)^2-4 \sigma^2\right) \Gamma\left(\frac{1}{2}\mp\lambda -\sigma \right) \Gamma\left(\frac{1}{2}\mp\lambda +\sigma \right) \\
&\bigg] \frac{\total s}{2\pi i} \eqend{.}
\end{split}
\end{equation}
For the terms which still depend on $\epsilon$ (the last two lines), we can shift the contour over the poles at $s=-\frac{1}{2}\pm\lambda$ and $s=-\frac{3}{2}\pm\lambda$, and the remaining contour integral is again bounded and vanishes as $\epsilon \to 0$. After using standard $\Gamma$ and $\psi$ recurrence identities, these terms then give
\begin{equation}
\frac{\abs{\lambda}}{2} \eqend{.}
\end{equation}
For the $\epsilon$-independent terms, we simplify the integral first using $\Gamma$ identities and rearrange terms to obtain
\begin{equation}
\begin{split}
&- \frac{\pi}{4} \int_{\Re s = - (1+2\Re\sigma)/4} \frac{\cos(2\pi\lambda) + \cos(2\pi\sigma)}{\left[ \cos(2\pi s) + \cos(2 \pi \lambda) \right] \left[ \cos(2\pi s) - \cos(2\pi \sigma) \right]} \bigg[ \\
&\qquad- 2 i \sin(2\pi\lambda) - \left( \mathe^{2i \pi \lambda} + \mathe^{-2i \pi s} \right) \frac{(1-2\lambda)^2 - 4 \sigma^2}{(3-2\lambda+2s) (1-2\lambda+2s)} \\
&\qquad+ \left( \mathe^{-2i \pi \lambda} + \mathe^{-2i \pi s} \right) \frac{(1+2\lambda)^2 - 4 \sigma^2}{(3+2\lambda+2s) (1+2\lambda+2s)} \bigg] \frac{\total s}{2\pi i} \eqend{.}
\end{split}
\end{equation}
In the last two terms, we perform a partial fraction decomposition
\begin{equation}
\frac{1}{(3-2\lambda+2s) (1-2\lambda+2s)} = \frac{1}{2 (1-2\lambda+2s)} - \frac{1}{2 (3-2\lambda+2s)} \eqend{,}
\end{equation}
shift the integration variable $s \to s-1$ in the second term and get
\begin{equation}
\begin{split}
&+ i \frac{\pi}{2} \int_{\Re s = - (1+2\Re\sigma)/4} \frac{\sin(2\pi\lambda) \left[ \cos(2\pi\lambda) + \cos(2\pi\sigma) \right]}{\left[ \cos(2\pi s) + \cos(2 \pi \lambda) \right] \left[ \cos(2\pi s) - \cos(2\pi \sigma) \right]} \frac{\total s}{2\pi i} \\
&\quad- \frac{\pi}{8} \left( \int_{\Re s = - (1+2\Re\sigma)/4} - \int_{\Re s = (3-2\Re\sigma)/4} \right) \frac{\cos(2\pi\lambda) + \cos(2\pi\sigma)}{\left[ \cos(2\pi s) + \cos(2 \pi \lambda) \right] \left[ \cos(2\pi s) - \cos(2\pi \sigma) \right]}  \\
&\qquad\bigg[- \left( \mathe^{2i \pi \lambda} + \mathe^{-2i \pi s} \right) \frac{(1-2\lambda)^2 - 4 \sigma^2}{(1-2\lambda+2s)} + \left( \mathe^{-2i \pi \lambda} + \mathe^{-2i \pi s} \right) \frac{(1+2\lambda)^2 - 4 \sigma^2}{(1+2\lambda+2s)} \bigg] \frac{\total s}{2\pi i} \eqend{.} \\
\end{split}
\end{equation}
The integral in the second line is given by the (negative) sum of the residues at $s = \pm\sigma$ and $s = \frac{1}{2}\pm\lambda$, which reads
\begin{equation}
- \frac{i}{2} \frac{\sin(2\pi\lambda)}{\sin(2 \pi \sigma)} \sigma \eqend{.}
\end{equation}
The integral in the first line can be done directly. We have
\begin{equation}
\begin{split}
&i \frac{\pi}{2} \int_{\Re s = - (1+2\Re\sigma)/4} \frac{\sin(2\pi\lambda) \left[ \cos(2\pi\lambda) + \cos(2\pi\sigma) \right]}{\left[ \cos(2\pi s) + \cos(2 \pi \lambda) \right] \left[ \cos(2\pi s) - \cos(2\pi \sigma) \right]} \frac{\total s}{2\pi i} \\
&\quad= \frac{1}{8\pi} \bigg[ \ln \left( \frac{\cos(\pi(s+\lambda))}{\cos(\pi(s-\lambda))} \right) + \frac{\sin(2\pi\lambda)}{\sin(2\pi\sigma)} \ln \left( \frac{\sin(\pi(s+\sigma))}{\sin(\pi(s-\sigma))} \right) \bigg]_{- (1+2\Re\sigma)/4 - i \infty}^{- (1+2\Re\sigma)/4 + i \infty} \\
&\quad= \frac{\abs{\lambda}}{2} -\frac{i}{2} \frac{\sin(2\pi\lambda)}{\sin(2\pi\sigma)} \sigma \eqend{,}
\end{split}
\end{equation}
so that the sum of all is
\begin{equation}
K = \abs{\lambda} - i \frac{\sin(2\pi\lambda)}{\sin(2 \pi \sigma)} \sigma
\end{equation}
and thus
\begin{equation}
J = \frac{e H}{\pi} \frac{\sinh(2\pi\abs{\lambda})}{\sin(2 \pi \sigma)} \sigma \eqend{.}
\end{equation}

\section{Particle detector} \label{detector}

Let us consider a particle interaction of the form
\begin{equation}
L_{\rm int} = -g (\phi^\dagger \psi \chi + h.c.),
\end{equation}
where $\psi$ is a (charged) detector particle, and $\chi$ is the product of the interaction. Following \cite{GKSSV}, the amplitude of interaction for a detector particle to 
annihilate with a $\phi$ antiparticle in a pair to give a neutral $\chi$ particle is given by:
\begin{equation}
A= \int \phi^*_{-k}\ \psi_q\ \chi^*_{q+k}\ g(t) a(t) dt,
\end{equation}
where, for convenience, we have allowed for a time dependent coupling in case we need to turn the interaction on and off. A convenient window function for this switching process is given by
\begin{equation}
g(t) = g \ a^{-\epsilon} e^{-\tilde \epsilon/a}.
\end{equation}
Here $\epsilon>0$ will play the role of an ultraviolet regulator for the momentum distribution, cutting off the interaction at late times, whereas $\tilde\epsilon>0$ is an infrared regulator, suppressing  the interaction at early times.

We shall restrict attention to the case $E=0$, for which $\lambda=0$, and the mode functions are given by 
\begin{equation}
\phi^*_{-k} = \phi^*_{k}= (2k)^{-1/2} W_{0,\sigma} (-2ik\eta),
\end{equation}
where, without loss of generality, we are assuming $k>0$. The Whittaker function can be written in terms of the integral representation
\begin{equation}
W_{0,\sigma} = {z^{\sigma+{1\over 2}} e^{-z/2} \over \Gamma\left({\sigma+{1\over 2}}\right)} \int_0^\infty e^{-zu} u^{\sigma-{1\over 2}} (1+u)^{\sigma-{1\over 2}} du. \label{intrepre}
\end{equation}
Further, we shall assume that the $\psi$ and $\chi$ particles are superheavy, so that, during the time when the interaction is switched on, their mode functions can be safely approximated by 
\begin{equation}
\psi_q \approx (2 w_\psi a)^{-1/2} e^{-i w_\psi t},\quad \quad \chi^*_{q+k} \approx (2 w_\chi a)^{-1/2} e^{+i w_\chi t}.
\end{equation}
Introducing $x=1/a$, the amplitude can be written as
\begin{equation}
A= {H^{-1}\over (8 k\ w_\psi w_\chi)^{1/2}} \int_0^\infty W_{0,\sigma} (2ik x/H) x^{-i {\Delta w\over H} -1 +\epsilon} e^{-\tilde \epsilon x} dx, \label{amplitude}
\end{equation}
where we have defined $\Delta w \equiv w _\chi - w_\psi \approx m_\chi-m_\psi$.
Substituting  (\ref{intrepre}) in (\ref{amplitude}) and doing the $x$ integration, we obtain
\begin{eqnarray}
A= {H^{-1}\over (8 k\ w_\psi w_\chi)^{1/2}} &&\left({2ik\over H}\right)^{\sigma+{1\over 2}} {\Gamma\left(\sigma-i {H^{-1}\Delta w} +{1\over 2} +\epsilon\right) \over \Gamma\left(\sigma+{1\over 2}\right)} \times \\
&& \int_0^{\infty} \left[\tilde\epsilon+{ik\over H}(1+2u)\right]^{-\left(\sigma-i{H^{-1}\Delta w} +{1\over 2}+\epsilon\right)} u^{\sigma-{1\over 2}} (1+u)^{\sigma-{1\over 2}} du.\nonumber
\end{eqnarray}
Removing the IR cut-off, $\tilde \epsilon\to 0$, the expression simplifies to 
\begin{eqnarray}
A_\sigma(\Delta w)= A_{-\sigma}(\Delta w) = {H^{-1} 2^{\sigma+{1\over 2}} \over (8 k\ w_\psi w_\chi)^{1/2}} & &e^{-{\pi \Delta w\over 2H}} i^{-\epsilon} \left({k\over H}\right)^{i{\Delta w\over H} -\epsilon} 
\times  \label{sigmasym} \\
{\Gamma\left(\sigma-i {H^{-1}\Delta w} +{1\over 2} +\epsilon\right) \over \Gamma\left(\sigma+{1\over 2}\right)} &&
 \int_0^{\infty} (1+2u)^{-\left(\sigma-i{H^{-1}\Delta w} +{1\over 2}+\epsilon\right)} u^{\sigma-{1\over 2}} (1+u)^{\sigma-{1\over 2}} du,\nonumber
\end{eqnarray}
where we have added the subindex $\sigma$ to the amplitude for later reference. The symmetry under the change $\sigma\to -\sigma$ is due to the analogous property of the Whittaker function $W_{\lambda,\sigma}$. The momentum dependence is now only in the prefactor, so that the probability of interaction is given by
\begin{equation}
{ dP} = {1\over 2\pi} |A_\sigma|^2 \propto k^{-(1+2\epsilon)} dk.
\end{equation}
If we remove the UV cut-off $\epsilon\to 0$, the distribution is logarithmic in $k$. This is to be expected, since it just expresses the fact that the probability of interaction is linear in time, and happens around fixed values $p_I$
of physical momentum 
\begin{equation}
p=k/a=p_I,
\end{equation}
so that $dk/k = Hdt$. However, this does not inform us about the typical value $p_I$ of this momentum at the time of interaction. 
For $\Delta w \geq m_\phi$, one might expect that this value should be dictated by kinematics:
\begin{equation}
p_I ^2 \approx (\Delta w)^2 - m_\phi^2,\label{expectation}
\end{equation}
although this is not necessarily the case, since the time of interaction could be dominated by virtual processes rather than by actual particles.

It is straightforward to check from (\ref{sigmasym}) that 
\begin{equation}
A^*_\sigma(\Delta w) = e^{-\pi {\Delta w\over H}} A_{-\sigma} (-\Delta w) = e^{-\pi {\Delta w\over H}} A_{\sigma} (-\Delta w),
\end{equation}
and therefore the transition probabilities satisfiy detailed balance at the Gibbons-Hawking temperature
\begin{equation}
dP(\Delta w) = e^{-2\pi \Delta w/H} dP(-\Delta w),
\end{equation}
as they should. However, this thermal character does not tell us which is the typical value of the momentum, $p_I$, which is likely to excite the detector.
Assuming the kinematic relation (\ref{expectation}), we may infer $p_I$ by analyzing the dependence of the amplitudes on $\Delta w$.

Using the relation between the Whittaker functions and the Bessel functions 
\begin{equation}
W_{0,\sigma}(z) = \sqrt{z\over \pi} K_\sigma(z/2),
\end{equation}
the amplitude in the limit $\tilde \epsilon \to 0$ can be calculated as
\begin{eqnarray}
A_\sigma &=&  {H^{-1} \over (8 k\ w_\psi w_\chi)^{1/2}} \left({2ik\over \pi H}\right)^{1/2} \int_0^\infty K_\sigma\left({ikx\over H}\right) x^{-iH^{-1}\Delta w +\epsilon-1/2} dx \nonumber \\
&=& {2^{-i {\Delta w\over H} + \epsilon} i^{-\epsilon}\over 4H(2 \pi  k\ w_\psi w_\chi)^{1/2}} \left({k\over H}\right)^{i{\Delta w\over H}-\epsilon} e^{-\pi \Delta w\over 2 H}  \times \nonumber \\
&&\Gamma\left( {1+2\epsilon\over 4}-{i\Delta w + H\sigma \over 2H}\right) \Gamma\left( {1+2\epsilon\over 4}-{i\Delta w - H\sigma \over 2H}\right).
\end{eqnarray}
In the limit $\epsilon \to 0$ this leads to 
\begin{equation}
\left|A_\sigma\right|^2 ={e^{-{\pi \Delta w\over H}} \over 32\pi H^2w_\psi w_\chi }  \left|\Gamma\left({1\over 4} -{i\Delta w + H\sigma \over 2H}\right)\Gamma\left({1\over 4} -{i\Delta w - H\sigma \over 2H}\right)
\right|^2{1\over k}.
\end{equation}
We are interested in the limit $m_\phi \gg 1$, where $\sigma \approx i m_\phi$. In this case, for $|m_\phi \pm \Delta w|\gg H$, we have
\begin{equation}
\left|A_\sigma\right|^2 \approx {\pi e^{-{\pi \Delta w\over H}} \over 4 Hw_\psi w_\chi \sqrt{|m_\phi^2-(\Delta w)^2|}} e^{-\pi{|m_\phi+\Delta w|+|m_\phi-\Delta w| \over 2H}}{1\over k}.\label{amplitudestirling}
\end{equation}  
Here, we have used 
\begin{equation}
\left|\Gamma\left({1\over 4} + i y\right)\right|^2 \approx 2\pi e^{-\pi|y|} |y|^{-1/2}.\quad\quad (y\in \mathbb{R}, |y|\gg 1).
\end{equation}
Thus, the leading dependence of the amplitude in $\Delta w$ is of the form
\begin{equation}
\left|A\right|^2 \propto \exp\left[-{\pi\over H} (\Delta w + \max  \{ m_\phi,|\Delta w|\})\right], \label{asq}
\end{equation}
where we have ignored the subleading denominator in (\ref{amplitudestirling}).

Note that the ``detector" will ``click" even if $m_\phi \gg \Delta w=m_\chi-m_\psi$. In this case it is perhaps not very appropriate to speak of a detection of a $\phi$ particle. Rather, the magnitude of the Boltzmann suppression in (\ref{asq}) suggests that the $\psi$ particle interacts with a fluctuation of the $\phi$ field, absorbing the energy $\Delta w$ sufficient for transforming the $\psi$ into a $\chi$ particle, and producing a $\phi$ antiparticle of mass $m_\phi$. If this is the case, Eq. (\ref{expectation}) will not be realized. On the other hand, if the mass $m_\phi$ is smaller than the gap $\Delta w$, it seems reasonable to expect that the momentum of the $\phi$ particle will be given by (\ref{expectation}). In this case, Eq. (\ref{asq}) suggests that the rate of detection of non-relativistic particles is much higher than that of relativistic ones. 



\end{document}